\documentclass{aa}

\usepackage{graphicx}
\usepackage{txfonts}
\usepackage{natbib}
\usepackage{siunitx}
\usepackage{multirow}
\usepackage{arydshln}
\usepackage{amsmath}

\usepackage{bm}
\newcommand{\orcid}[1]{\unskip\protect\href{https://orcid.org/#1}{\protect\includegraphics[width=8pt,clip]{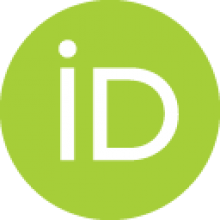}}}

\usepackage[colorlinks=true,allcolors=blue]{hyperref}

\makeatletter
\renewcommand*\aa@pageof{, page \thepage{} of \pageref*{LastPage}}
\makeatother

\begin{document}

\title{Sub-Jovian desert of exoplanets at its boundaries}
\subtitle{Parameter dependence along the main sequence}

   \author
        {Gy. M. Szab\'o          
            \inst{1,2,3} \orcid{0000-0002-0606-7930}
          \and
        Sz. K\'alm\'an
        \inst{2,4,5,6,7} \orcid{0000-0003-3754-7889}
            \and
           L. Borsato \orcid{0000-0003-0066-9268}
          \inst{8}  
          \and
         V. Heged\H{u}s \orcid{0000-0001-7699-1902}
         \inst{3,5} 
         \and
        Sz. M\'esz\'aros \inst{1,3}
         \and
         R. Szab\'o \inst{4,7,9,10} \orcid{0000-0002-3258-1909}
          }

\institute{ 
    ELTE E{\"o}tv{\"o}s Lor\'and University, Gothard Astrophysical Observatory, Szombathely, Szent Imre h. u. 112., H-9700, Hungary
    \and 
     MTA-ELTE Exoplanet Research Group, Szombathely, Szent Imre h. u. 112., H-9700, Hungary
    \and
    MTA-ELTE  Lend{\"u}let "Momentum" Milky Way Research Group, Hungary
    \and
    Konkoly Observatory, Research Centre for Astronomy and Earth Sciences,  ELKH, Budapest, Konkoly-Thege Miklós út 15–17., H-1121, Hungary
\and
    ELTE E{\"o}tv{\"o}s Lor\'and University, Doctoral School of Physics,  Budapest, Pázmány Péter sétány 1/A, H-1117, Hungary
\and 
    Graduate School of Physics, University of Szeged, Szeged, D\'om t\'er 9., H-6720, Hungary
\and  
    CSFK, MTA Centre of Excellence, Budapest, Konkoly Thege Miklós út 15-17., H-1121, Hungary \\
     \email{xilard1@gothard.hu}
\and 
   INAF-Osservatorio Astronomico di Padova Vicolo dell'Osservatorio 5, I-35122, Padova, Italy
 \and
    E\"otv\"os Lor\'and University, Institute of Physics, P\'azm\'any P\'eter s\'et\'any 1/A, H-1171 Budapest, Hungary
\and
    MTA CSFK Lend\"ulet Near-Field Cosmology Research Group
}

\abstract%
    {The lack of sub-Jovian planets on orbits of $P_{\rm orb} < 3$~days is a puzzling aspect  of galaxy formation with regard to the distribution of exoplanets whose origins are currently unresolved.}
    {The possible explanations behind the formation of the sub-Jovian or Neptunian desert include several scenarios that can lead to different shapes for the boundary, predicting various dependencies between the position of the boundary and the stellar parameters.}
    {We explored the exoplanet distribution in various 2D and 3D projections, revealing the stellar-dependent substructures in the $P_{\rm orb}$--$M_P$ and the $P_{\rm orb}$--$R_P$ parameter plane.}
    {We demonstrate that the upper boundary includes a range of planets, namely, inflated hot Jupiters and normal hot Jupiters, in the two parameter planes, respectively. We confirm the dependence of the boundary on several stellar parameters and, based on a fuzzy clustering analysis, we provide quantitative formulae for the dependencies in groups of smaller and larger planets. The overall period-radius distribution shows chemical substructures as well, with the boundary being dependent on volatiles and alpha-elements, alongside marginal (to none) dependence found for refractory elements.}{These findings confirm multiple plausible causes for the formation of the desert,  particularly preferring those scenarios related to the irradiation-driven loss of the atmospheres of moderately massive planets as the predominant process in shaping planetary distributions.}
        
   \date{Recieved ....; accepted ....}

   \keywords{Methods: statistical --
                  Planets and satellites: formation -- Astronomical databases: miscellaneous}
\maketitle
\section{Introduction}

The observed absence of short-period planets ($\lesssim 3$~d) below the hot Jupiter
clump and above the hot super-Earths is known
as the ``sub-Jovian desert'' \citep{2011ApJ...727L..44S,2011A&A...528A...2B,2014ApJ...787...47S,2015MNRAS.452.3001C,2016ApJ...820L...8M,2017AJ....153..130E,2018MNRAS.479.5012O,2019MNRAS.485L.116S} and ``Neptunian desert'' \citep{2016A&A...589A..75M,2018A&A...610A..63D,2018MNRAS.476.5639I,2022AJ....163..298M}. In recent years, several planets have been observed in these desert regions \citep{2018A&A...610A..63D,2020ApJ...903L...6D,2020NatAs...4.1148J,2020Natur.583...39A,2021A&A...653A..60M,2022AJ....163..298M}, with some deserts possibly undergoing a conversion into a ``savanna.''  In this paper, we describe the desert-savanna boundaries in the parameter space.

There is a known planet population both above and below its boundaries, therefore, the sub-Jovian or Neptunian desert or savanna itself is a puzzling structure of planetary distribution. The location of the desert is a meeting point for high-energy physics and atmospheric  non-local thermodynamic equilibrium (NLTE) processes (representing the stellar activity and irradiation), planetary thermodynamics, and magneto-hydrodynamical interactions between the planet and the star, end-points of planetary migration, accretion processes in general, and planet formation and evolution in extreme environments as well. Therefore, a test for planet formation theories is to check how they can explain the presence of the desert and its boundaries.

\begin{figure}
    \centering
    \includegraphics[width = 0.48\textwidth]{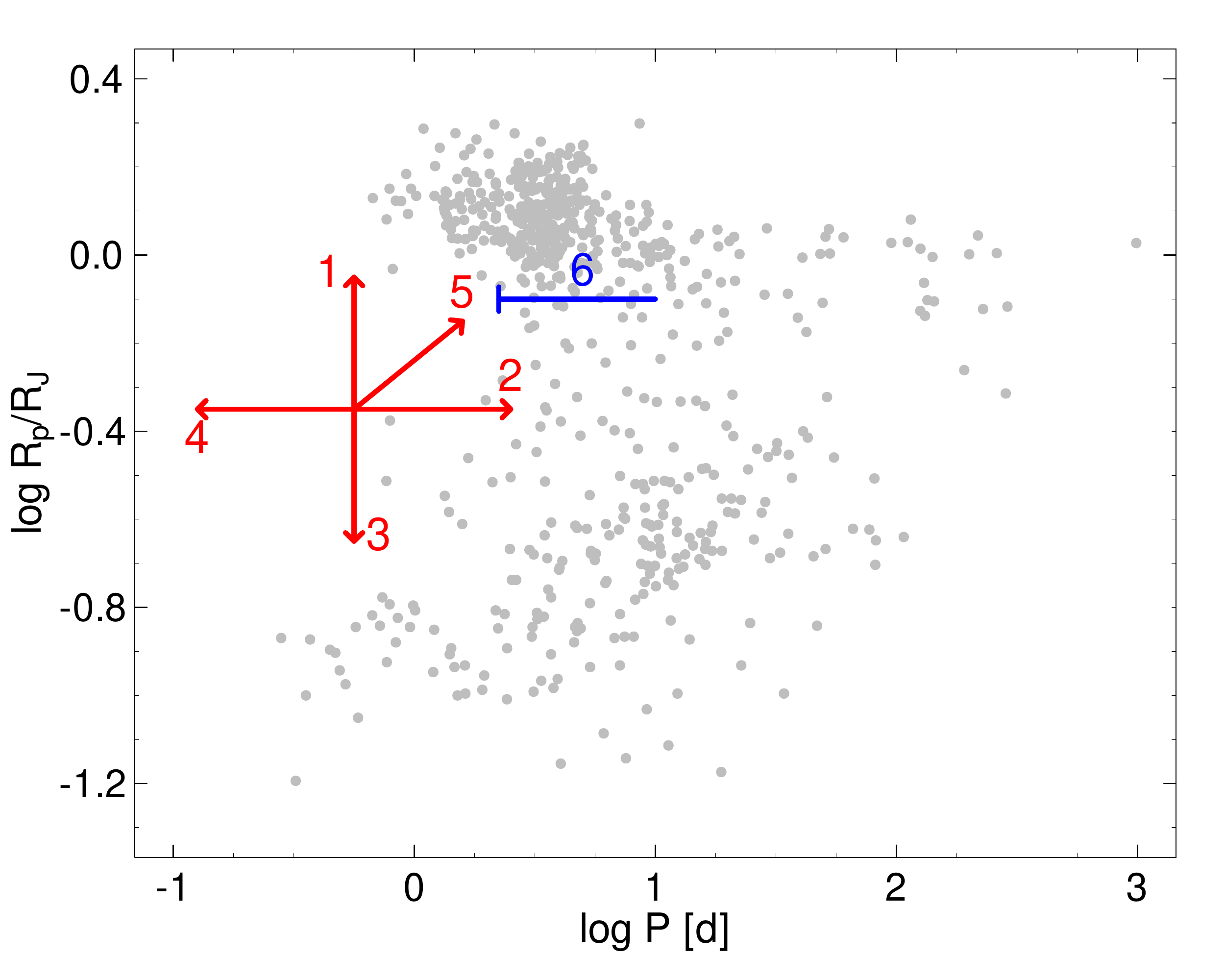}
    \caption{Sketch of possible paths for already formed planets leaving the sub-Jovian desert (red arrows) and planets not allowed to reach the desert (blue arrows) overplotted to the distribution of known exoplanets. Theories and references behind the scenarios are summarized in the text.}
    \label{fig:theories}
\end{figure}

In Fig.\ref{fig:theories}, we summarize the possible paths of how a young planet, once formed, can leave the desert. The indicative direction of ``leaving the desert'' is denoted by numbers in the figure. The processes suggested for the different scenarios thus far include:
1) the hyperinflation of low-mass gas giants at the boundary \citep{2015IJAsB..14..201M}; 2) planet-growth-based processes, namely,\ it is has been well established that hot Jupiters have higher-than-expected (i.e., inflated) radii \citep{2018AJ....155..214T}, which can be attributed to irradiation from the host \citep[e.g.,][]{2021A&A...645A..79S}. As the desert is observed in both the  $R_P$--$P$ and $M_P$ -- $P$ planes (and at sub-Jovian planet sizes), these processes would necessarily have to include accretion as well as inflation. The theoretical calculations done so far \citep{2021A&A...656A..69E, 2021A&A...656A..70E} were insufficient  to explore this scenario in detail; 3) high-eccentricity migration \citep{2017AJ....154..192G,2018MNRAS.479.5012O} suggests that the orbits of close-in planets with initially high eccentricities circularize with higher orbital periods; 4) decreasing planet size can be the result of photoevaporation \citep{2014ApJ...792....1L,2016NatCo...711201L, 2018MNRAS.479.5012O}, Roche-lobe-overflow \citep{2014ApJ...783...54K}, ``boil-off'' \citep{2016ApJ...817..107O}, and impact-based mass-loss \citep{2015Icar..247...81S, 2018SSRv..214...34S}.
Finally, (5) the tidal migration of hot Neptunes \citep{1986ApJ...309..846L, 2022ApJ...931...10R},
tidally trapped outward migration \citep{2006ApJ...642..478M}, and disk migration \citep{2016A&A...589A..75M, 2021A&A...648A..69A} are other alternative scenarios. 

Besides scenarios involving migration and evaporation, \cite{2018ApJ...866L...2B} found that an accretion parameter correlated with planet mass and disk inner edge can explain the upper right side of the desert and explains its suggested slope of $-2/7$. \cite{2021A&A...648A..69A} suggested a further possibility, where orbital resonances change the orbital periods of the inner planets. The curious case of TOI-2196 b \citep{2022A&A...666A.184P} would fit in well with migration-based scenarios, being a small and volatile-rich planet present in the savanna.

\cite{2016A&A...589A..75M} derived relationships for the lower and upper boundaries of the desert in the $R_P$--$P$ plane finding that at the upper edge, $R_P \propto P^{-\frac{1}{3}}$, and at the lower edge,  $R_P \propto P^{\frac{2}{3}}$, thus suggesting different origins in the two size-regimes.
\cite{2016ApJ...820L...8M} explained the two boundaries as the tidal disruption barrier for gas giants following their high-eccentricity migrations and because of different mass-radius laws of low-mass and high-mass exoplanets, the two edges are formed differently.
\cite{2018MNRAS.479.5012O} suggested that the photoevaporation and the high-eccentricity migration to form the upper and lower boundaries, respectively. \cite{2019AREPS..47...67O} summarizes the processes that could have formed the desert boundaries from an evaporation approach, with a very detailed introduction to the astrophysical processes.
%Attributing the upper boundary to tidal disruption of giant planets undergoing high-eccentricity migration \cite{2016ApJ...820L...8M} has been considered as an alternative candidate with more precise predictions. 
A somewhat similar structure in the period-radius distribution is related to the observed increasing sub-Saturn ($\approx$~4--8~R$_\oplus$) occurrence rate up to $\sim$300 days orbital period. \cite{2022ApJ...924....9H} argued for a radiation-induced process (resulting in atmoshperic mass-loss), which is efficient at transforming sub-Saturns into sub-Neptunes ($\lesssim$4 R$_\oplus$), with short orbital periods. The short-period end of this process is similar to the ``process 3'' in our Fig.\ref{fig:theories} and could explain the
lower boundary of the desert as well. 

In our earlier paper \cite{2019MNRAS.485L.116S}, we found that the boundaries of the desert depend more on fundamental stellar parameters and less on planetary mass, density, and size. In that work, the sample of exoplanets were smaller, which limited the further inference. In this paper, we re-apply the same analyses to the larger exoplanet data set at hand and we include a few new tests as well. We show that the boundaries of the desert in the period-mass ($P$ -- $M_P$) and the period-radius ($P$ -- $R_P$) planes are marked by a different planet population. The desert in the two main projections can be considered as a sign of multiple scenarios of planetary evolution acting differently in the various parameter spaces. 
Therefore, we revisited the $P$--$M_P$ and $P$--$R_P$ boundaries separately, which allowed for the findings of \cite{2019MNRAS.485L.116S} to be quantified. 
The dependence of the $P$--$M_P$ and $P$--$R_P$ boundaries on the stellar parameters are to be considered as another piece of evidence to support this assertion. Finally, we give expressions for both the planetary radius and mass in terms of stellar parameters and via fuzzy clustering. We conclude that in all projections, planets form two main groups that behave differently, in a possible connection with the unresolved processes behind the formation of the sub-Jovian or Neptunian desert.

\section{Methods and data selection}\label{sec:meth}

%We undertook the analysis of the distribution of exoplanets around the so-called Neptune desert of exoplanets in the planetary radius -- orbital period and planetary mass -- orbital period parameter spaces, simultaneously. 
\subsection{Data on  confirmed planets from the NASA exoplanet archive}

The input to this analysis consists of two data sets. We followed \cite{2019MNRAS.485L.116S} in building up the sample from the NASA Exoplanet Archive (5009 planets in total\footnote{The data were obtained from the Planetary Systems Composite Data database \url{https://exoplanetarchive.ipac.caltech.edu/cgi-bin/TblView/nph-tblView?app=ExoTbls&config=PSCompPars} on 5 April 2022.}), following the same filtering as in \cite{2019MNRAS.485L.116S}. This filtering has left out planets with less constrained parameters and kept the ``golden sample'' only. In building up the test sample, we defined the following selection criteria: (i) planetary mass $M_P < 13 M_J$ and density $\rho_P < 25\rho_J$ and (ii) relative uncertainty for the planetary radius and mass of $<20\%$ and $<60\%$, respectively, leaving us with a sample of $650$ exoplanets. This is an increment with regard to the 406 planets in our previous analysis \citep{2019MNRAS.485L.116S} and our current sample also represents a better quality because it also contains the filtering for the mass determination errors. We accepted the stellar parameters provided in the original NASA Exoplanet Archive records. 

In Fig. \ref{fig:theories}, we show the distribution of our filtered sample (of $650$ exoplanets) in the $R_P$--$P$ plane. Figures \ref{fig:rho}--\ref{fig:meta} show the same sample in both the radius-period and mass-period parameter spaces.
%In Fig. \ref{fig:theories}, we show the distributions of stellar and planet parameters in the $M_P$--$P$ and $R_P$--$P$ planes. In Figs. \ref{fig:rho}--\ref{fig:meta} the same scatterplots are shown with a coloring third parameter, where the colors represent the value of the examined parameter. 
This design enables the visualization of the exact values of the continous variable, but has the limitation of being merely qualitative. To be able to quantitatively see the structures in the sample, we applied two-sampled Kolmogorov-Smirnov (KS) tests \citep[see e.g.,][]{feigelson_babu_2012} to the projected parameters in different regions (univariate bands) of the data distribution.

Still following \cite{2019MNRAS.485L.116S}, we defined two statistical regions in the $R_P$ -- $P$ plane as: the area between $0.28$ and $0.63\ R_J$ is labeled $A$, and the one between $0.06$ and  $0.16 R_J$ is labeled $B$. In the $M_P$ -- $P$ plane, we defined the $C$ region between $0.032 - 0.305\ M_J$ and the $D$  region between $0.001 - 0.010\ M_J$. 

To observe whether a stellar or planetary parameter is selective at the position of the border (in other words, the position of the border depends on the examined parameter), we compared the period distribution of two subsamples in $A$ to those in $B$; and two subsamples in $C$ to those in $D$. The subsamples were separated at the median of the examined parameter. For example, we compared the period distribution of high-temperature and low-temperature stars in $A$ and $B$, etc. We compared the resulting period distributions in a two-sampled KS test in all four regions and computed the $p$ values. If the resulting $p$ is small, we can conclude that the two compared samples are drawn from different distributions; or, in other words, the third parameter can affect the distribution of exoplanets. Because regions $A$ and $C$ are in the desert and $B$ and $D$ are outside, this comparison helps identify those parameters that affect the formation and the boundaries of the desert.

Following the technique of colored scatter plots, we sometimes identified several groups of exoplanets following a specific pattern. These features can be proven with clustering methods. Here, we followed the fuzzy clustering algorithm of \cite{bezdek},  as implemented in the \verb|FKM| function in the \texttt{R} software package\footnote{R Core Team (2017). R: A language and environment for statistical computing. R Foundation for Statistical Computing,
Vienna, Austria. \url{https://www.R-project.org/}}.

\subsection{APOGEE catalog of planet-host stars}\label{sec:apogeepl}

The second database we used for the analyses is based on Apache Point Observatory Galactic Evolution Experiment \citep[APOGEE,][]{majewski_2017}, which provides homogeneously derived atmospheric stellar parameters and eliminates uncertainties originating from the usage of different model atmospheres, spectral synthesis codes, line lists, and so on. We built an APOGEE exoplanet host stars catalog to test the findings on this homogeneous data set and we plot the data in Figs. \ref{fig:plan-apo}-\ref{fig:metacomp}.

As part of the fourth iteration of Sloan Digital Sky Survey \citep[SDSS-IV, ][]{blanton_2017}, in its final data release \citep[DR17, ][]{abdurro_2021}, the APOGEE-2 survey derived and reported main atmospheric parameters and individual element abundances for 733,901 stars across the entire sky \citep{zasowski_2017} and focused on observing as many planet-host stars as possible, so it is an excellent choice to cross-match our sample with that of DR17. Observations were made with the identical SDSS spectrographs \citep{wilson_2019} mounted on the 2.5-meter Sloan Foundation Telescope \citep{gunn_2006} and the 2.5-meter Irénée du Pont Telescope of Las Campanas Observatory, Chile. The resolving power is $R\sim22,500$ and the spectral coverage of the near-infrared H-band -- from 15140~\AA~to 16940~\AA. This makes it possible to select planet-host stars near the Milky Way disk and dust obscured regions of the Milky Way. 
Besides observing stellar targets, APOGEE is sensitive to substellar mass companions (exoplanet candidates) too, as multi-epoch visits along with high-precision RV measurements are available \citep{majewski_2017}.

The APOGEE DR17 collaboration published both raw and calibrated values of effective temperature, surface gravity, and applied offset corrections to $\rm [M/H]$ based on solar neighborhood stars. Here, we take the calibrated effective temperatures and overall metallicities into account as $T_{\rm eff}$ is in a better agreement with the infrared flux method (IRFM) scale than the raw spectroscopic values \citep{abdurro_2021}. For this analysis, we carefully select the most accurate and precise APOGEE data. We globally filter out APOGEE stars labeled by the quality control flag \texttt{STAR\_BAD}\footnote{\url{https://www.sdss.org/dr17/irspec/apogee-bitmasks/} \\The web documentation  include detailed flagging descriptions.} \citep{abdurro_2021} (but retained \texttt{STAR\_WARN}), then we cross-matched the exoplanet catalog with the APOGEE DR17 data set in \texttt{TOPCAT} version 4.8-2 \citep[Tool for OPerations on Catalogues And Tables, ][]{topcat_2005} based on the equatorial coordinates of the host stars ($\Delta<1 \rm ~arcsec$). The next step involved the implementation of several quality criteria: relative error lower than $20\%$ for $\rm R_{\rm P}$, $\texttt{VSCATTER}<1$~km/s for RV variations, $\texttt{VERR}<1$~km/s for RV error to filter out potential variable stars, and $\rm \texttt{M\_H\_ERR}<0.1$~dex for the metallicity uncertainty.

We note that the selected elements fall into the APOGEE data reduction categories of ``most reliable'' and ``reliable'' with regards to the overall quailty of their derivation (see the APOGEE DR17 website for more details).

\section{Results} \label{sec:res_desert}

%\subsection{Multiform boundaries in the parameter space}
%Fig 2.
\subsection{Different planets on the borders in projections}\label{sec:rho}
\begin{figure}
    \centering
    \includegraphics[width = .48\textwidth]{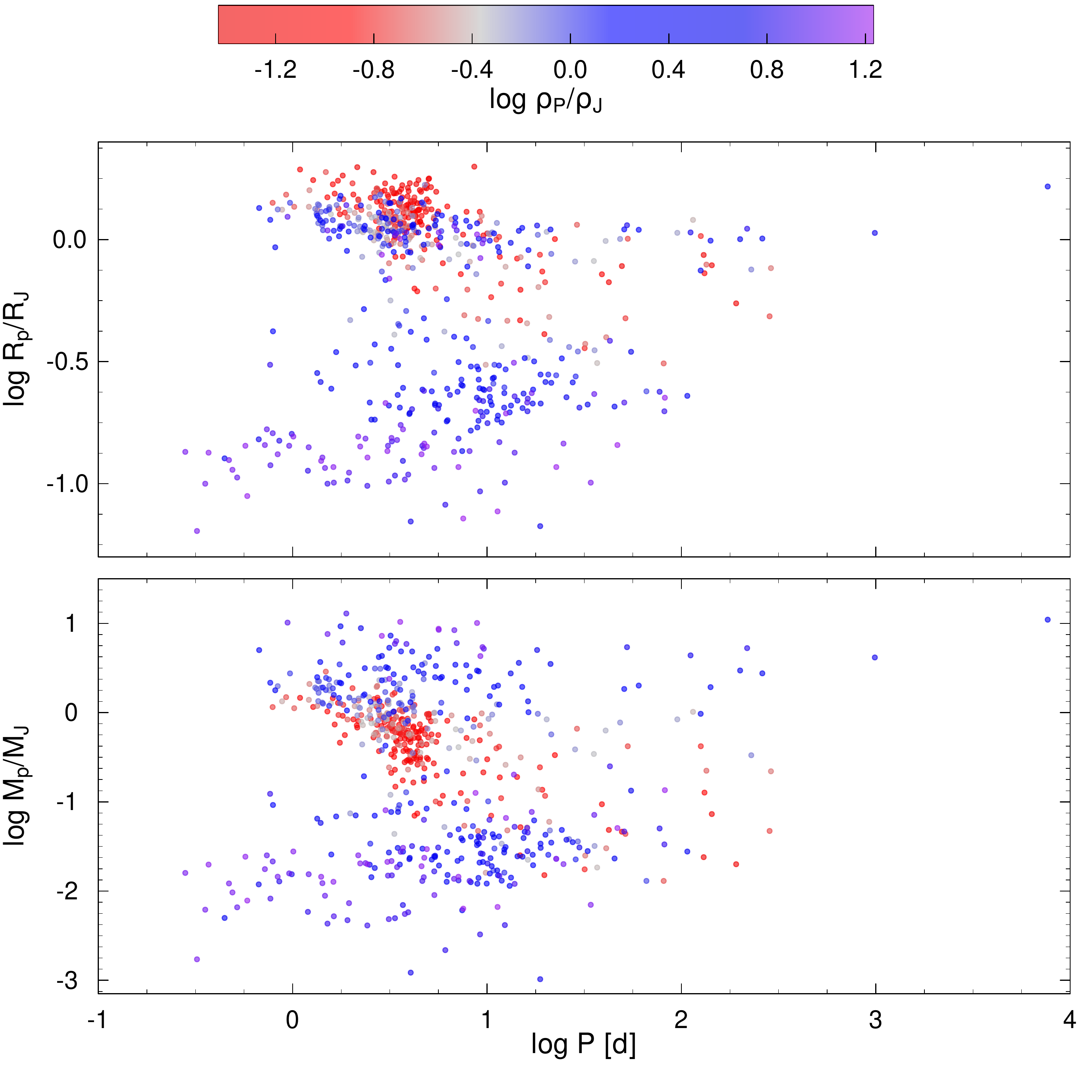}
    \caption{Sub-Jovian/Neptune desert of exoplanets in the $R_P$ -- $P$ and $M_P$ -- $P$ parameter spaces (top and bottom panel), colored by the planetary density. In the top panel, the upper boundary is marked by ``normal'' Jupiters with higher density ($> 1 \rho_J$), while in the bottom panel, the population of inflated planets (red points) appear below the normal Jupiters and, thus, the inflated Jupiters mark the actual boundary.} 
    \label{fig:rho}
\end{figure}

In Fig. \ref{fig:rho}, we show the filtered NASA Exoplanet Archive planets in the $R_P$--$P$ and $M_P$--$P$ planes with the planet's mean density as a coloring variable. The comparison of the two scatter plots reveals an interesting feature. Planets with $\gtrsim \rho_J$ density, plotted as blue dots, follow a similar distribution in both planes, with the most important difference of the compressed size range of hot Jupiters (upper panel) that is related to well-understood planetary physics. However, the group of low-density planets (plotted in red) behave very differently between the two plots. These planets have $\lesssim M_J$ mass and they can be found at the middle of the mass distribution (lower panel), but they are among the most inflated planets, and can be observed on the top of hot Jupiters in the radius distribution (upper panel).

This distinct clump of planets suffers from a very significant dislocation compared to the non-inflated or moderately inflated planets, which requires explanation. Also, the low-density planets themselves  form the upper boundary of the sub-Jovian (or Neptunian) desert in the $M_P$--$P$ projection, while they tend to be farther from the boundary than the hot Jupiters in the $R_P$--$P$ plane. Indeed, in the $R_P$--$P$ plane, the hot Jupiters themselves form the upper boundary of the desert.

The anonymous referee has pointed out that
similar structures could be simple byproducts of the well-known increasing bulk density above a 1.0 $M_J$ planetary mass, due to the increasing self-compression of the gas that builds up the planetary body. To test the robustness of this finding, we used an empirical model to describe the mass-density relations in general and we removed this contribution from the scatter plot. 

\begin{figure}
    \centering
    \includegraphics[width = .48\textwidth]{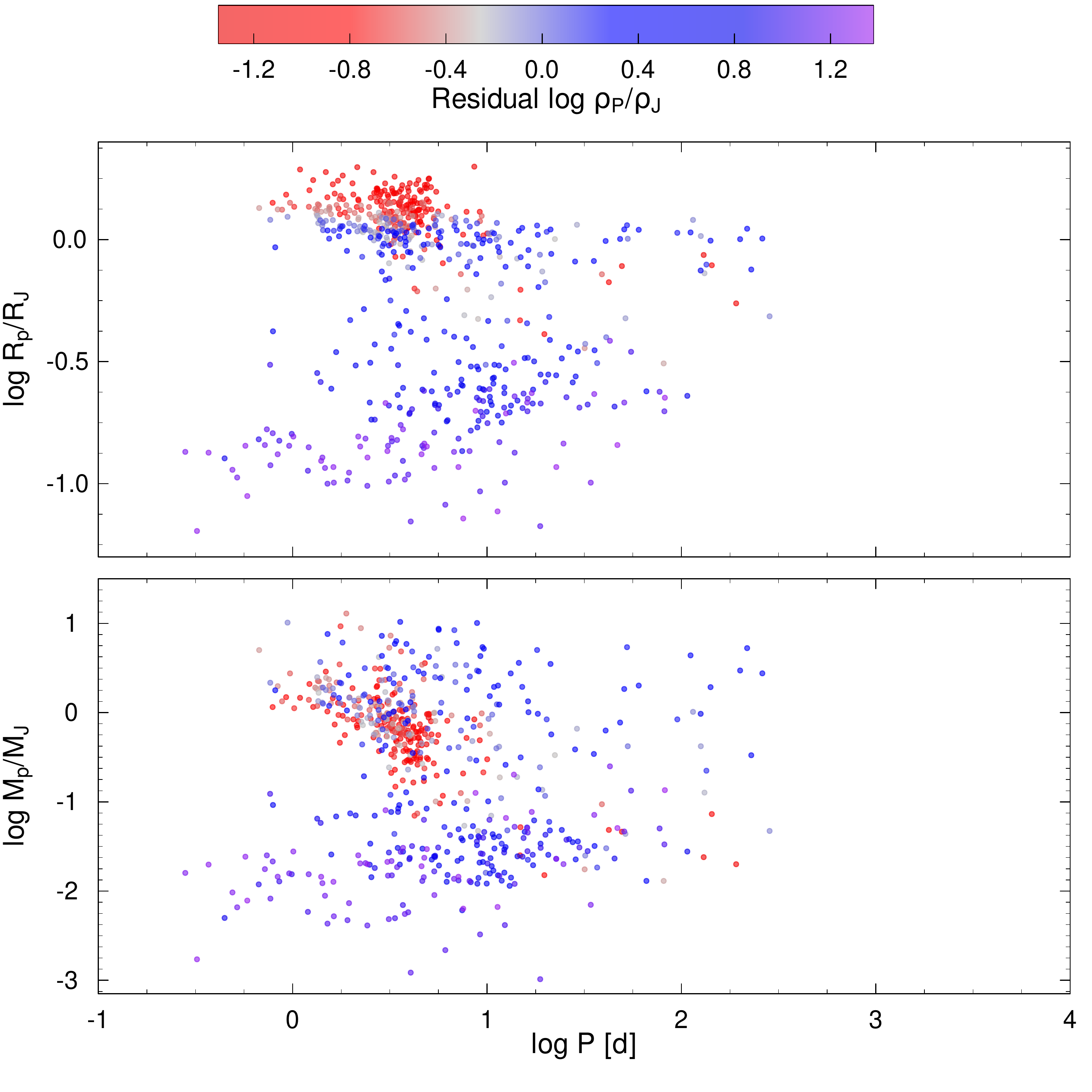}
    \caption{Scatterplots of the distribution of exoplanets in the $R_P$ -- $P$ and $M_P$ -- $P$ planes as in Fig. \ref{fig:rho} but with synthetic densities subtracted from the observed ones (see text for details).} 
    \label{fig:synthrho}
\end{figure}

We calculated synthetic densities for each planet, following \cite{2011ESS.....2.2403T, 2012A&A...547A.112M} and based on mean radii (Eq. (24) of \citep{2012A&A...547A.112M}. Since the applied model equation is designed to reproduce theoretical radii for planets, with semi-major axis $a < 0.1$ AU and $0.1$ AU $< a < 1$ AU, we further restricted our sample to include planets with $a < 1$ AU. Subtracting the logarithm of the resulting densities yielded the distribution shown in Fig. \ref{fig:synthrho}. We find that the upper edge of the desert in the $M_P$ -- $P$ plane (bottom panel of Fig. \ref{fig:synthrho}) is still generally dominated by low residual-density planets, which are then then observed to be further away from the sub-Jovian desert in the $R_P$ -- $P$ plane (top panel of Fig. \ref{fig:synthrho}). 

This suggests that the general density-mass relation does not explain the observed phenomenon. Indeed, the scatter plots colored with the residual $\log$ densities (Fig. \ref{fig:synthrho}) show a very similar distribution to what we see in Fig. \ref{fig:rho} and we can have the subjective visual impression that Fig. \ref{fig:synthrho} is even clearer. In any case, our conclusion is that the found double population of close-in giants is not a consequence of the density-mass relations, but it is rather superimposed onto the general pattern of density distributions. Taking the position of the planets with $\rho/\rho_J<0$ into account, it seems to be quite plausible that their presence is primordially influenced by the desert boundaries.
We also point out that the exoplanets with longer orbital periods ($>10$ days) which have lower densities (red points on Fig. \ref{fig:rho}) remain roughly in the same position relative to the desert in both planes of Fig. \ref{fig:rho}.

The analysis of the bulk densities (Fig. \ref{fig:rho}) thus suggests the upper boundary of the desert has a multiform nature: it is marked by ordinary hot Jupiters in one projection and inflated hot Jupiters in the other projection. Since the planetary physics behind these two distinct classes of hot Jupiters are different, this also brings up the possibility that the formation of the desert is a result of multiple processes (at least at the upper boundary): one process acting on ordinary jot Jupiters and one another acting on inflated hot Jupiters.

The possible biases behind the observed distribution can be excluded by simple quantitative considerations. In particular, $137$ exoplanets falling into the $\rho_P < 0.35\ \rho_J$ category were discovered primarily with $18$ different photometric facilities including both ground-based and space telescopes, so their distribution in the parameter-space cannot be devoted to some instrument-specific selection distortion of some specific exoplanet discovery surveys. Also, from the entire sample of $650$ exoplanets, only $23$ of them were not discovered with the transit method (mostly radial velocity planets that had later been observed in transit as well) and none of them belong\ to the inflated group of planets (according to the definition given at the beginning of this paragraph). Therefore, the presence of radial velocity discoveries should not be considered as sources of biased data. Out of the $137$ exoplanets with $\rho_P < 0.35\ \rho_J$, an overwhelming majority ($116$) are members of planetary systems with a single known planetary mass object, with $19$ systems containing 2 planets and $2$  containing $2$ and $3$ planets. The brightness distribution of host stars in the two samples is also very similar, with $V = 10.56 \pm 1.94$\;mag for the larger, $650$-member sample and $10.89 \pm 1.59$\;mag for the low-density sample. We therefore see no evidence for observational biases. 
We can consider two possible explanations for the multiple planet populations at the upper boundary: (i) either the inflated planets are in the early stages of planetary evolution and they are still in the process of initial contraction or losing their atmospheres; or (ii) they are more mature planets, but have a different inner structure than the higher density hot Jupiters. In either case, the processes that form the desert must be selective for these formation scenarios, leading to the two distinct planet populations at the two different projections of the upper boundary.

To test these scenarios, a wider sample with more precisely determined exoplanet parameters is necessary. Our understanding of these questions will be greatly improved by the two upcoming ESA missions: Ariel \citep{2021arXiv210404824T} and PLATO \citep{2014ExA....38..249R}.

%We consider how the forseeable results of Ariel \citep{2021arXiv210404824T}, or the very precise and accurate detection of more exoplanets by  PLATO \citep{2014ExA....38..249R} will contribute to the understanding of these questions in the second part of this paper series (K\'alm\'an et al., submitted, hereafter Paper II). 

\subsection{Dependence of the boundary on stellar parameters}

In this subsection, we analyze the $R_P$--$P$ and $M_P$--$P$ scatter plots with regard to the distribution of stellar parameters (effective temperature, stellar radius, stellar mass, $\log g$, and metallicity), complementing and extending the results described in \cite{2019MNRAS.485L.116S}.

\subsubsection{Effective temperature of the host}\label{sec:teff}

\begin{figure}
    \centering
    \includegraphics[width = 0.48\textwidth]{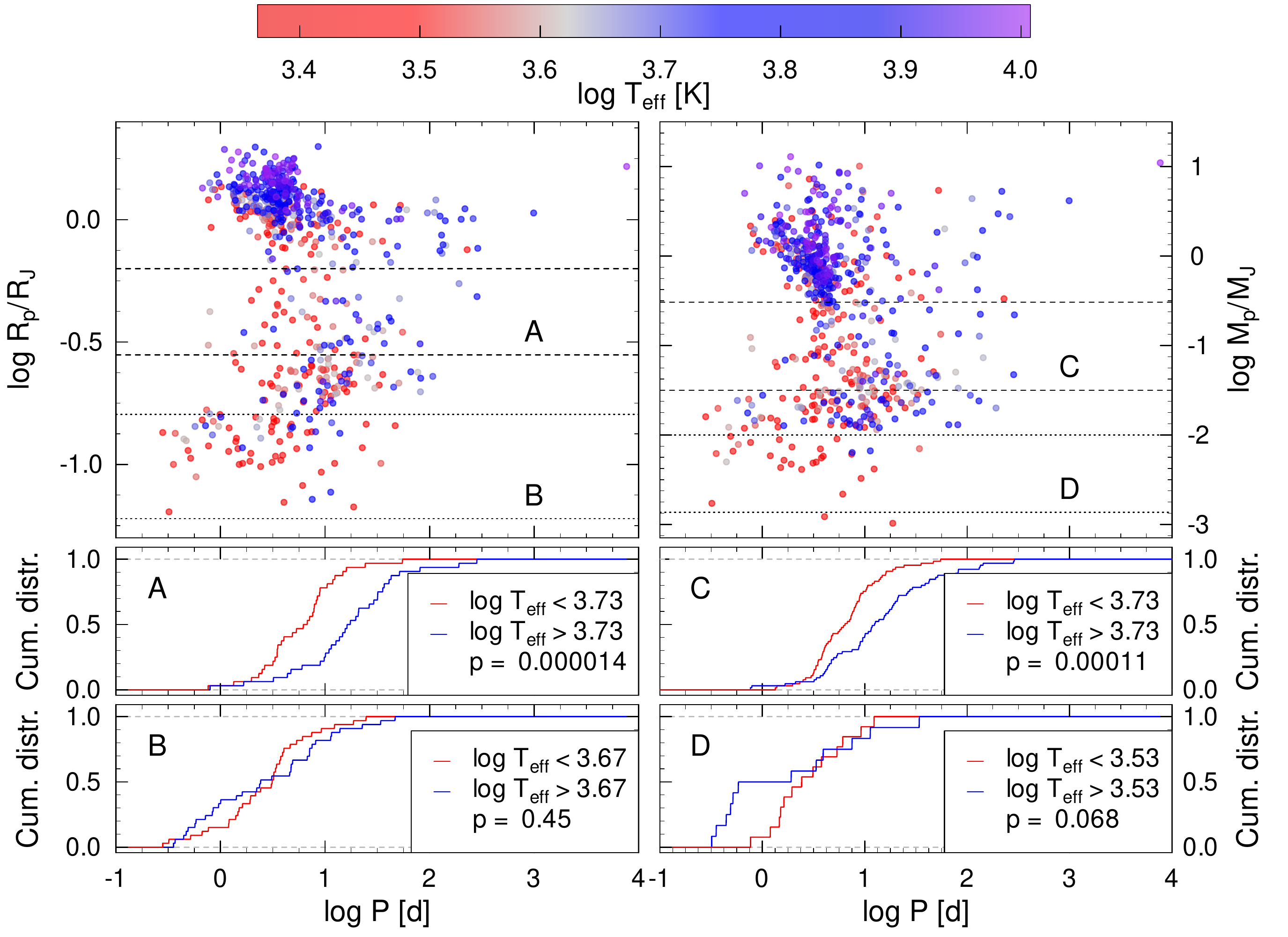}
    \caption{Scatterplot showing the distribution of exoplanets in the $R_P$ -- $P$ (top left) and $M_P$ -- $P$ (top right) parameter spaces, colored by the temperature of the host star (see the color bar for the values). The middle and bottom panels show the cumulative distribution of the two samples divided at the median of $T_{\rm eff}$ in the $A$, $B$, $C,$ and $D$ regions. The $p$-values of the KS-test are also shown.}
    \label{fig:teff}
\end{figure}

Following the recipe described in Sect. \ref{sec:meth}, we split the sample of exoplanets at the median of the effective temperature of the host stars in different regions separately. The split values were $5370$\;K, $4677$\;K, $5370$\;K, and $3388$\;K for the $A$, $B$, $C,$ and $D$ regions, respectively (see Fig. \ref{fig:teff}). We found that in both parameter spaces, planets in $A$ and $C$ regions tend to have shorter orbital periods around cooler hosts. The $p$-values of the KS tests were: $10^{-5}$ and $10^{-4}$ in $A$ and $C$, respectively. This discrepancy in the cumulative distribution has not been observed in the $B$ or $D$ regions ($p$-values: $0.45$ and $0.07$), proving that the observed dependence is specifically located at the right boundary of the desert. Therefore, we confirm the findings of \cite{2019MNRAS.485L.116S} that the effective temperature of the host star plays a key role in the formation of the desert and it does so in both parameter spaces.

\subsubsection{Stellar radius} \label{sec:rad}

\begin{figure}
    \centering
    \includegraphics[width = 0.48\textwidth]{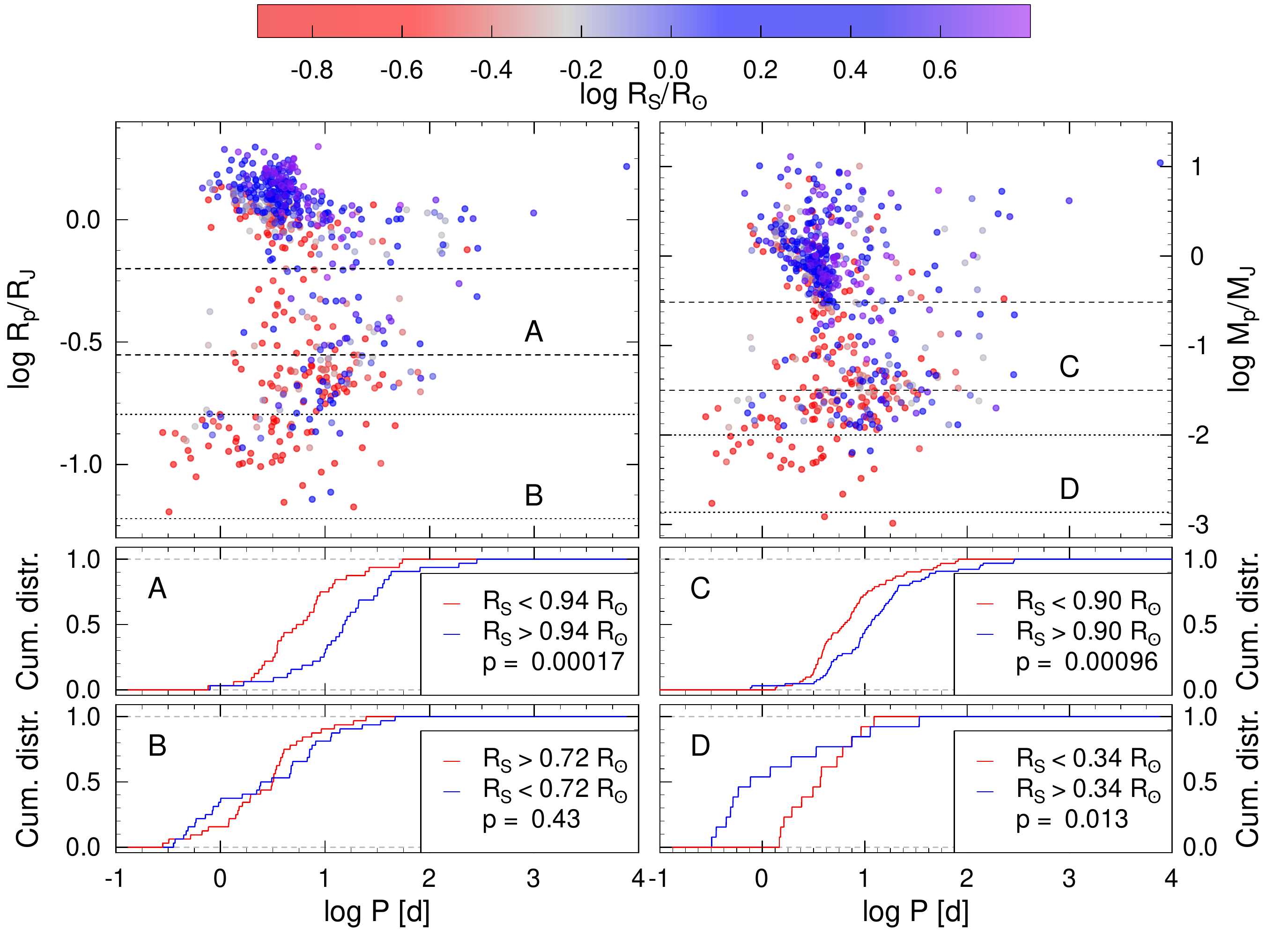}
    \caption{Scatterplots showing the distribution of exoplanets in the $R_P$ -- $P$ and $M_P$ -- $P$ parameter spaces with $R_S$ used as the third parameter (top row) and the respective cumulative distributions (middle and bottom rows; same as Fig. \ref{fig:teff}).%     showing the distribution of exoplanets in the $R_P$ -- $P$ (top left) and $M_P$ -- $P$ (top right) parameter spaces (same as Fig. \ref{fig:teff}), but with $R_S$ used as the third parameter instead of $T_{\rm eff}$.
    }
    \label{fig:rad}
\end{figure}

The boundary of the desert also shows dependence on $R_S$ in both the $R_P$ -- $P$ and $M_P$ -- $P$ planes, as visible in the scatter plots of Fig. \ref{fig:rad}. The stars below the median radii ($0.94\ R_{\sun}$ and $0.90\ R_{\sun}$ in the $A$ and $C$ regions, respectively) are more likely to host planets with shorter orbital periods.  We note that the KS test yields $p$-values of $2 \cdot 10^{-4}$ and $10^{-5}$ in $A$ and $C$, respectively. There is no significant difference in the cumulative distribution planets in the $B$ and $D$ regions. In $B$, the $p$-value is $0.43$; while in $D$, the $p=0.013$ still corresponds to $2.48\sigma$, proving the statistical insignificance of the (otherwise apparent) differences seen in the bottom right panel of \ref{fig:rad}. 
As we find that stars with lower radii tend to host planets with shorter orbital periods and that this phenomenon is not present in the control regions, we conclude that the stellar radius also plays a key role in the formation of the desert.

\subsubsection{Stellar mass}\label{sec:mass}

\begin{figure}
    \centering
    \includegraphics[width = 0.48\textwidth]{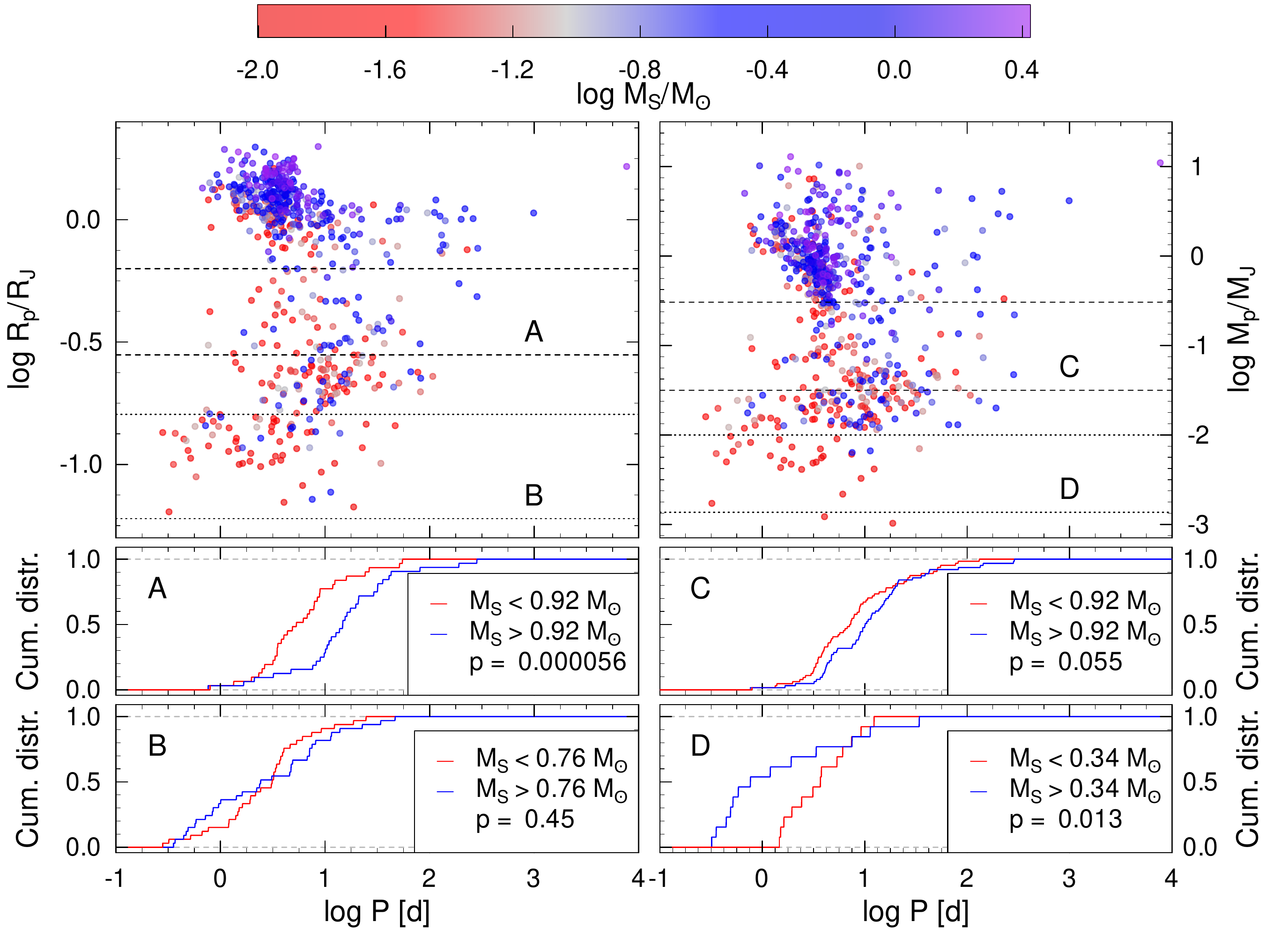}
    \caption{Scatterplots showing the distribution of exoplanets in the $R_P$ -- $P$ and $M_P$ -- $P$ parameter spaces with $M_S$ used as the third parameter (top row) and the respective cumulative distributions (middle and bottom rows; same as Fig. \ref{fig:teff}).}
    \label{fig:mass}
\end{figure}

The scatter plot of exoplanets with the mass of the host star used as the third parameter is shown in Fig. \ref{fig:mass}. The split values of $M_S$ are $0.92\ M_{\sun}$, $0.76\ M_{\sun}$, $0.92\ M_{\sun}$, and $0.34\ M_{\sun}$ in $A$, $B$, $C,$ and $D$, respectively. Dividing the samples in the four regions at these values leads to an ambiguous detection of $M_S$ dependence of the desert boundary (i.e., the distribution of planets at the size ranges of the desert). This is expressed in terms of the $p$-values of $6\cdot 10^{-5}$ opposed to $0.45$ in $A$ and $B$ regions, and then $0.055$ opposed to $0.013$ in $C$ and $D$ regions. The dependence is present in the $R_P$--$P$ plane and is insignificant in the $M_P$--$P$ plane, consistently with the results of \cite{2019MNRAS.485L.116S}. Therefore, we find that stellar mass has some influence most importantly on the radius distribution, but does not play an essential role in the formation of the desert on its own.

\subsubsection{Stellar $\log g$}

\begin{figure}
    \centering
    \includegraphics[width = 0.48\textwidth]{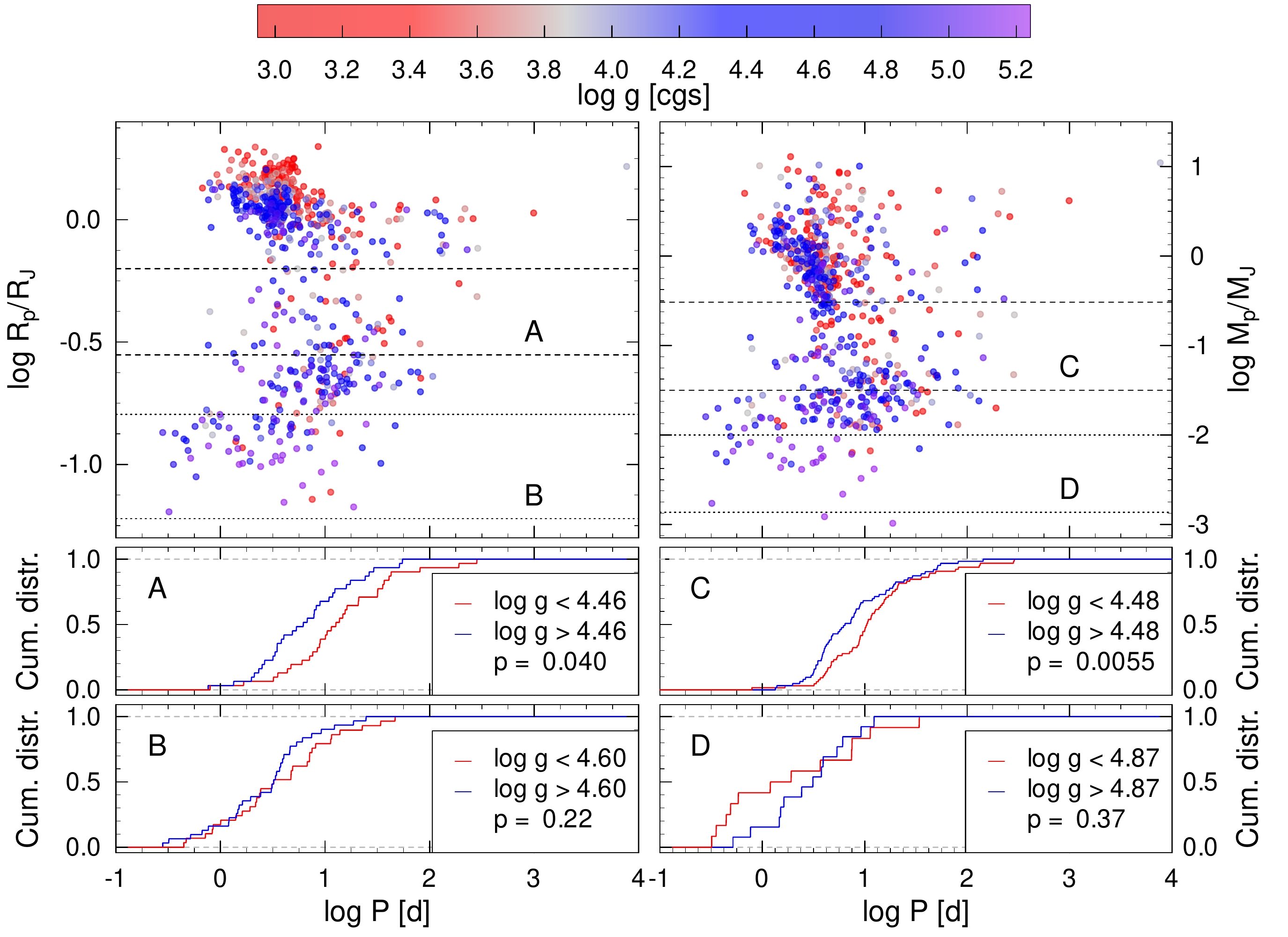}
    \caption{Same as Fig. \ref{fig:teff} but with $\log g$ used as the third parameter instead of $T_{\rm eff}$.}
    \label{fig:logg}
\end{figure}

Exploring the effects of $\log g$ on the distribution of exoplanets in the two parameter spaces, we divided the samples at $\log g = 4.46$ in the $A$ region, $\log g = 4.60$ in $B$, $\log g = 4.48$ in $C$, and $\log g = 4.87$ in $D$. We find that there is no statistical difference in the cumulative distributions in either the $A$ or $B$ regions of the $R_P$ -- $P$ plane. This is expressed by the $p$-values of the KS-test: $0.040$ in  $A$ (corresponding to $\sim 2.1 \sigma$) and $0.22$ in $B$.
In the size ranges of the desert of the $M_P$ -- $P$ parameter space, there is a hint of a discrepancy between the cumulative distributions of the sample divided at the median at $p = 0.0055$ (corresponding to $2.78 \sigma$). There is no such difference in the sample of the control region ($p = 0.34$). 

We therefore find no firm dependence of the desert on the surface gravity of the host star, contradicting to the earlier results of  \cite{2019MNRAS.485L.116S} (based on fewer planets and unconstrained for a reliable mass determination). The structure of Fig. \ref{fig:logg} the most inflated hot Jupiters are hosted by stars with the lowest $\log g$ values ($< 3.2$) from the sample.

\subsubsection{Stellar metallicity}\label{sec:meta}

\begin{figure}
    \centering
    \includegraphics[width = 0.48\textwidth]{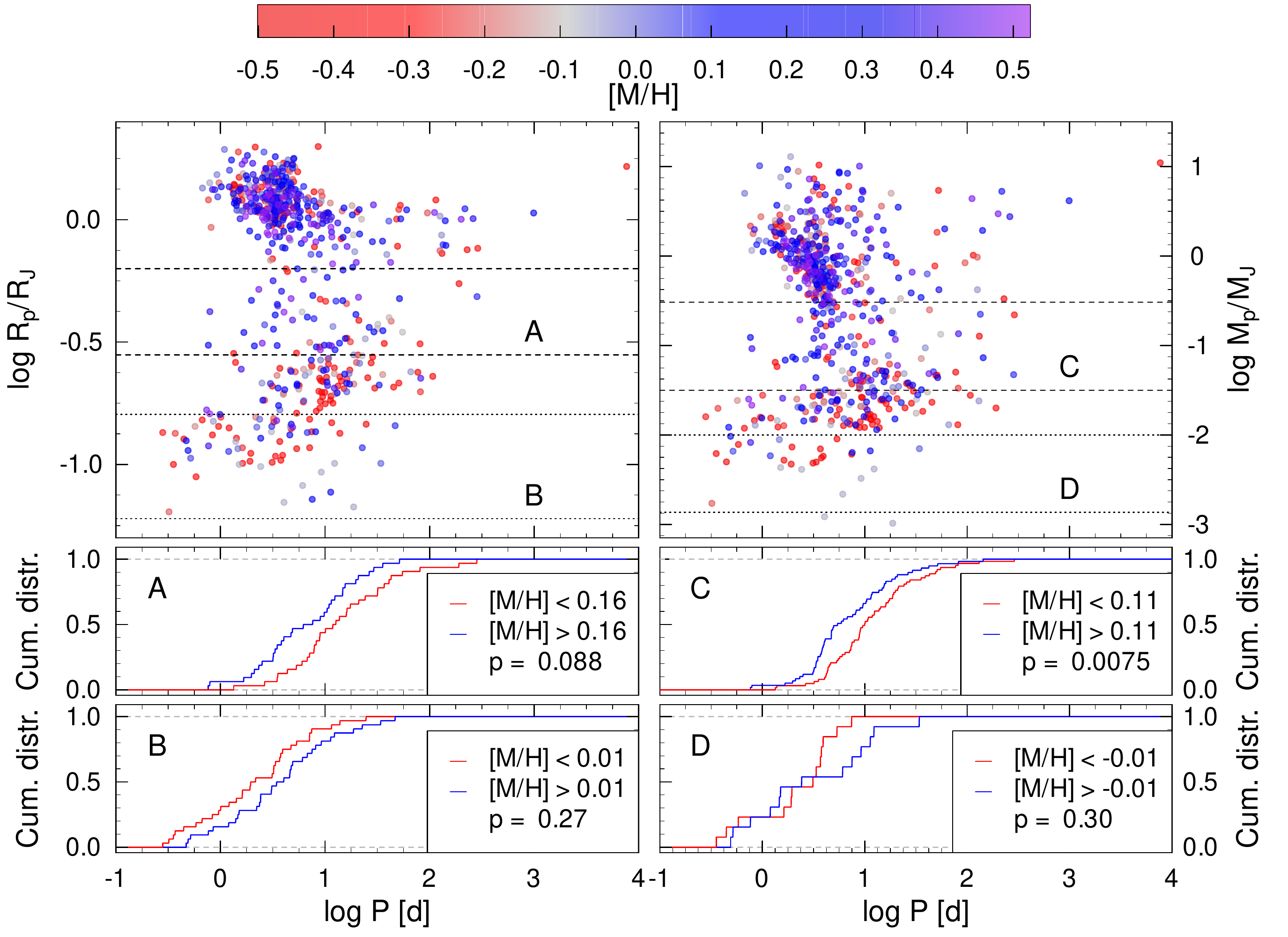}
    \caption{Same as Fig. \ref{fig:teff} but with [M/H] used as the third parameter instead of $T_{\rm eff}$.}
    \label{fig:meta}
\end{figure}

In the filtered sample of $650$ used in the exploration of the previous parameters, there are $10$ stars with unknown metallicities and we omitted those from the current analysis. The samples in $A$, $B$, $C,$ and $D$ are split at [M/H]$ = 0.16$, $0.01$, $0.11,$ and $-0.01$, respectively. There is no significant difference in the cumulative distribution of planets in the two examined areas of the $R_P$ -- $P$ plane ($p$-values are $0.088$ and $0.27$ in $A$ and $B$). In the case of the $M_P$ -- $P$ plane, there is also only a hint that planets from the $C$ region are more likely to have shorter orbital periods around younger, more metal-rich stars, expressed via $p = 0.0075$ (corresponding to $2.67 \sigma$) in $C$ compared to $0.30$ in $D$. 

As in the case of $\log g$, we find no firm dependence of the sub-Jovian desert of exoplanets on the metallicity of the host stars. This is somewhat contradictory to the results of \cite{dong, petigura} and \cite{2019MNRAS.485L.116S}.

%{\bf More discussion using APOGEE data?}

\subsection{Multimodal dependence of the planetary mass and radius on stellar parameters}

\begin{figure*}
    \centering
    \includegraphics[width = \textwidth]{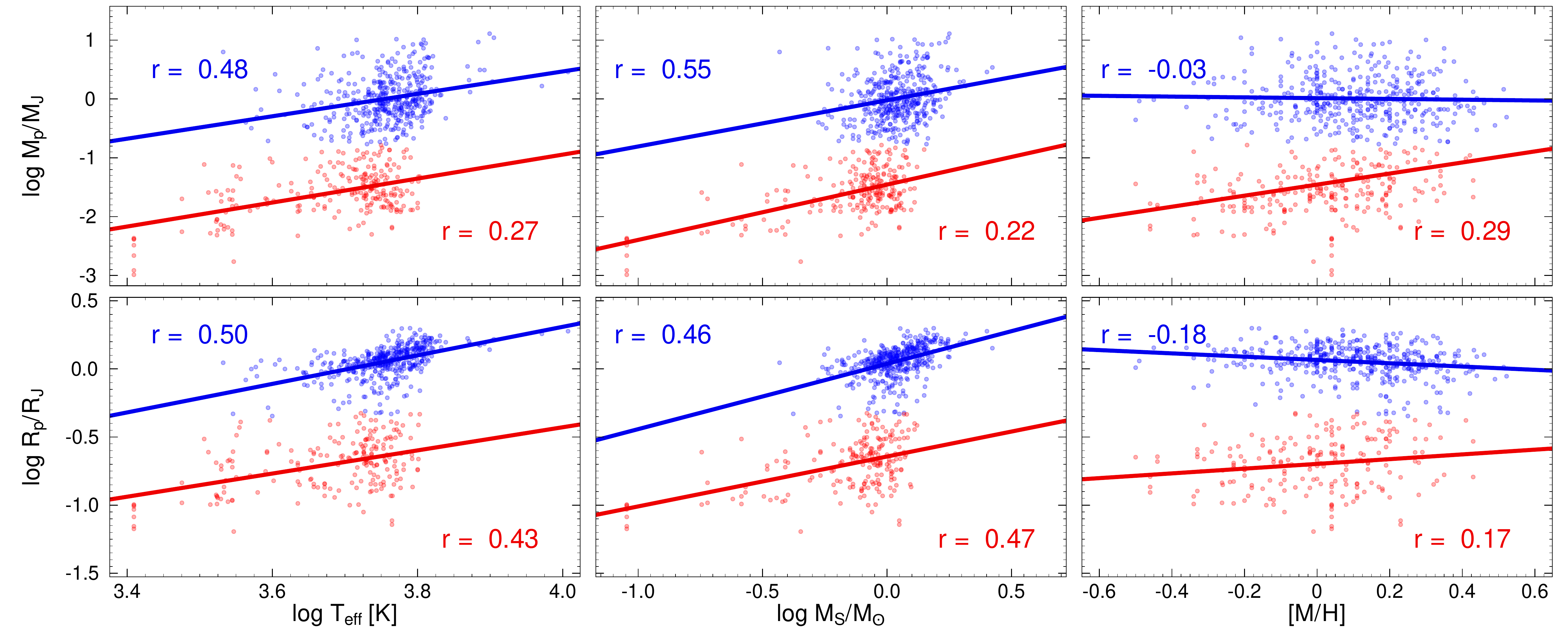}
    \caption{Correlation between planet mass (top row), planetary radius (bottom row), and the main stellar parameters: $T_{\rm eff}$ (left column), $M_S$ (middle column), and [M/H] (right column). The samples are divided into two groups via a fuzzy clustering algorithm applied to the distributions in the right panels. Blue points: Cluster of ``larger planets.'' Red points: Cluster of `smaller planets.''
    The fitted linear trends of Eqs. (\ref{eq:grt})--(\ref{eq:smfe}) are plotted with solid blue and red lines. Pearson's $r$ values, corresponding to the giant and smaller exoplanets are also displayed with blue and red, respectively.}
    \label{fig:deps}
\end{figure*}

The experienced bimodality of the planet groups at the border of the desert Fig. \ref{fig:teff} and the experienced differences behind the $R_P$--$P$ and $M_P$--$P$ relations reflect the presence of two different type of planets at the border: hot Jupiters at the upper boundary and super-Earths at the lower boundary. As the boundary of the desert in the mass and radius projections depend on stellar parameters, we expect to see various groups of planets with different parameter dependencies between the radius, mass, and stellar parameters.

To explore distinct correlation laws between the planet and stellar parameters, we divided our sample into two subgroups in both the mass-metallicity and the radius-metallicity planes via a $k$-means fuzzy clustering algorithm \citep{fclust}. In the radius-metallicity plane, the resulting cluster consisting hot Jupiters had a median radius of $1.15\ R_J$ (interquartile range of 1.01-1.33) and a cluster consisting Neptune- and Earth-sized planets had a median radius of $0.21\ R_J$ (interquartile range of 0.14--0.27). 
When the clustering was done in the mass-metallicity distribution, the cluster of large planets had a median mass of $0.93\ M_J$ (interquartile range of 0.57--1.80) and the subgroup of smaller planets has a mean mass of $0.026\ M_J$ (interquartile range of 0.015--0.048). 

We fit linear regression models describing the dependence of planet mass and radius on $T_{\rm eff}$, $M_S$, and [M/H]. Between the planet size and the effective temperature, we get:

%\begin{align}
 % \log \left( \frac{M_P}{M_J}\right)  & = 1.91(34) \cdot \log \left( \frac{T_{\rm eff}}{1\ \text{K}} \right) -7.15(33), & r = 0.48\label{eq:gmt}\\
  % \log \left( \frac{R_P}{R_J} \right)  & = 1.05(9) \cdot \log \left( \frac{T_{\rm eff}}{1\ \text{K}} \right) -3.88(33), & r = 0.50\label{eq:grt} 
%\end{align}
\begin{align}
 \log \left( \frac{M_P}{M_J}\right)  & = 1.91(34) \cdot \log \left( \frac{T_{\rm eff}}{1\ \text{K}} \right) -7.15(1.27), & r = 0.48,\label{eq:gmt}\\
 \log \left( \frac{R_P}{R_J} \right)  & = 2.05(9) \cdot \log \left( \frac{T_{\rm eff}}{1\ \text{K}} \right) -3.88(33), & r = 0.50,\label{eq:grt} 
\end{align}

in the ``large planet'' cluster and 
\begin{align}
   \log \left( \frac{M_P}{M_J}\right)  & = 2.04(25) \cdot \log \left( \frac{T_{\rm eff}}{1\ \text{K}} \right) -9.09(91), & r = 0.27, \label{eq:smt}\\
   \log \left( \frac{R_P}{R_J} \right)  & = 0.85(12) \cdot \log \left( \frac{T_{\rm eff}}{1\ \text{K}} \right) - 3.83(46), & r =0.43,  \label{eq:srt} 
\end{align}
in the ``small planets'' cluster. In the expressions, the ambiguity (standard deviation) of the last digits are shown in parentheses after the value of each coefficients. 

The expressions involving the stellar mass are:
%\begin{align}
 %  \log \left( \frac{M_P}{M_J}\right)  & = 0.78(17) \cdot \log \left( \frac{M_S}{M_\sun} \right) -0.02(2), &  r = 0.55 \label{eq:gmm}, \\
  % \log \left( \frac{R_P}{R_J} \right)  & =  0.48(4) \cdot \log \left( \frac{M_S}{M_\sun} \right) -0.04(5), &  r = 0.46 \label{eq:grm} 
%\end{align}
\begin{align}
   \log \left( \frac{M_P}{M_J}\right)  & = 0.78(17) \cdot \log \left( \frac{M_S}{M_\sun} \right) -0.02(2), &  r = 0.55 \label{eq:gmm}, \\
   \log \left( \frac{R_P}{R_J} \right)  & =  0.48(4) \cdot \log \left( \frac{M_S}{M_\sun} \right) -0.04(1), &  r = 0.46, \label{eq:grm} 
\end{align}

in the ``large planet'' cluster and 
\begin{align}
    \log \left( \frac{M_P}{M_J}\right)  & = 0.94(9) \cdot \log \left( \frac{M_S}{M_\sun} \right) -1.46(3), & r = 0.22 \label{eq:smm}, \\
   \log \left( \frac{R_P}{R_J} \right)  & = 0.37(5) \cdot \log \left( \frac{M_S}{M_\sun} \right) - 0.64(1), & r = 0.47, \label{eq:srm} 
\end{align}
in the ``small planet'' cluster. Such a correlation between planet size and stellar mass (but without clustering) was also noted by \cite{fortney, wu}, and \cite{2021A&A...652A.110L}. \cite{wu} found the linear coefficient between $\log M_P$ and $\log M_S$ to be unity, which is in good agreement with Eq. (\ref{eq:smm}). These findings are also consistent with the qualitative observations from Fig. \ref{fig:mass} that the most massive stars from our sample ($M_S > 1.5\ M_{\sun}$) host the largest and least dense hot Jupiters, those with radii  of $> 1.4\ R_J$.

For the stellar metallicity dependencies, we get
\begin{align} 
    \log \left( \frac{M_P}{M_J}\right)  &= -0.07(11) \cdot {\rm[M/H]}+0.01(2), & r = -0.03 \label{eq:gmfe}, \\
   \log \left( \frac{R_P}{R_J} \right)  &= -0.12(3) \cdot {\rm[M/H]} + 0.07(1), & r = -0.18, \label{eq:grfe} 
\end{align}
in the ``large planet'' cluster and 
\begin{align} 
    \log \left( \frac{M_P}{M_J}\right)  &= 0.94(9) \cdot {\rm[M/H]}- 1.45(3), & r = 0.29 \label{eq:smfe}, \\
   \log \left( \frac{R_P}{R_J} \right)  &= 0.17(7) \cdot {\rm[M/H]} - 0.70(7), & r = 0.17 \label{eq:srfe}  
\end{align}
in the ``small planet'' cluster.
\cite{wu} suggested that there is no correlation between stellar metallicity and planetary mass, which is compatible to the low values of correlation coefficients found in our analysis.

The $p$ value of these multiple correlations are in the range of $10^{-4}$ and $10^{-16}$ for all equations --except the ones involving [M/H] (Eqs. \ref{eq:gmfe}-\ref{eq:srfe}), which have $p$ values between 0.01--0.5, showing marginally significant or even, insignificant correlations. The bimodality of the correlations illustrate that there is no one-to-one relationship between planet sizes and stellar parameters. It can be said, however, that in almost every case (with the exception of $R_P$ -- $M_S$), a stronger correlation is found in the case of large exoplanets. In these linear models, no underlying astrophysical connections are considered, they are based solely on statistics; however, in future works, they can be used to further consider the reasons and processes which shape these dependencies.

\subsubsection{Elemental abundances from APOGEE}

The APOGEE stellar parameters were taken from a homogeneous, state-of-the-art quality catalog of stellar data, while the sources of stellar data in the NASA Exoplanet Archive are inhomogeneous, although the latter one includes many more planets, which can increase the significance of the statistical analysis based upon it.

\begin{figure*}
    \centering
    \includegraphics[width = \textwidth]{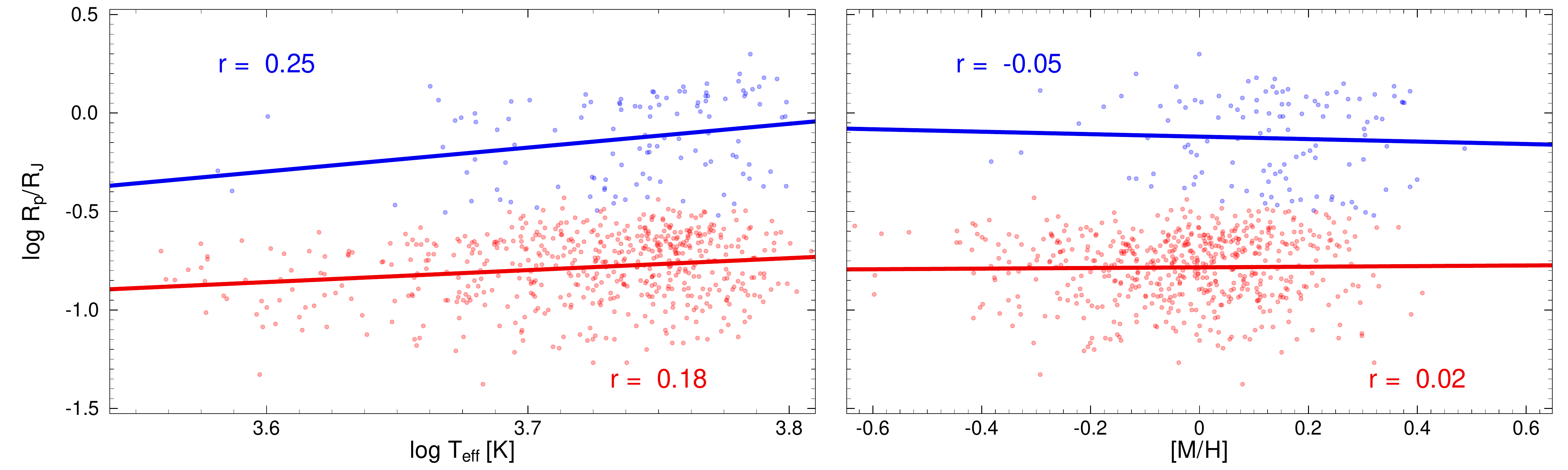}
    \caption{Correlations between $R_P$ -- $T_{\rm eff}$ (left panel) and $R_P$ -- [M/H] (same as the lower left and lower right panels of Fig. \ref{fig:deps}) showing the fuzzy clustering of APOGEE data and with the linear regressions of Eqs. (\ref{eq:apogrfe})--(\ref{eq:aposrfe}) overplotted.}
    \label{fig:plan-apo}
\end{figure*}

\begin{figure*}
    \centering
    \includegraphics[width = \textwidth]{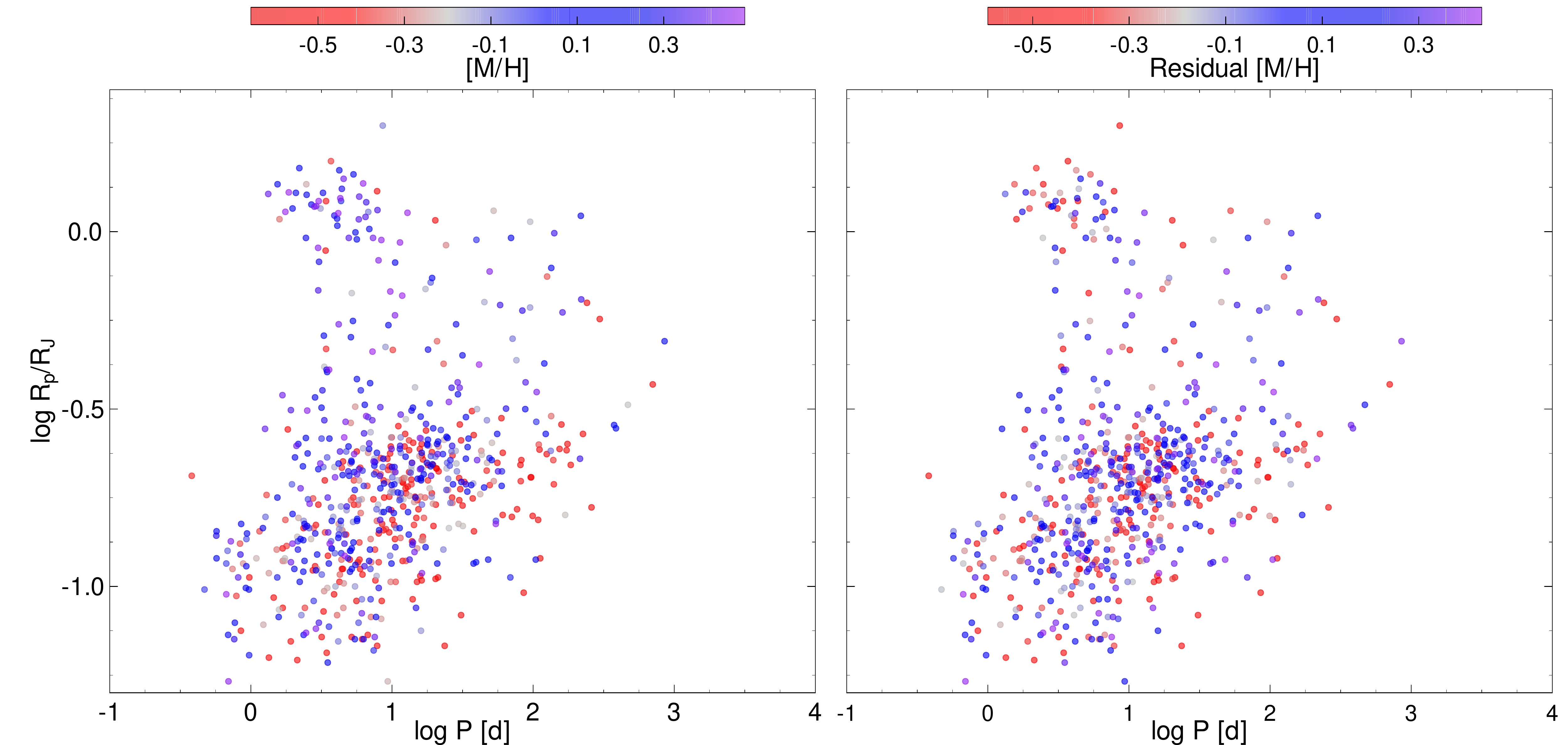}
    \caption{Distribution of exoplanets from the APOGEE planet host sample with [M/H] as the coloring parameter (left panel) and the residuals after a bilinear fit (right panel).}
    \label{fig:metacomp}
\end{figure*}

From the APOGEE planet sample (Sect. \ref{sec:apogeepl}), we repeated the fits described in Eqs. (\ref{eq:grt}), (\ref{eq:grm}), (\ref{eq:smt}), and (\ref{eq:smm}) leading to the following regressions:
\begin{align}
    \log \left( \frac{R_P}{R_J} \right)  & = 1.21(43) \cdot \log \left( \frac{T_{\rm eff}}{1\ \text{K}} \right) -4.65(1.60), & r = 0.25\label{eq:apogrt}\\
    \log \left( \frac{R_P}{R_J} \right)  &= -0.03(5) \cdot {\rm[M/H]} - 0.11(3), & r = -0.05 \label{eq:apogrfe}
\end{align}
 in the ``large planet'' cluster and
\begin{align}    
    \log \left( \frac{R_P}{R_J} \right)  & = 0.61(14) \cdot \log \left( \frac{T_{\rm eff}}{1\ \text{K}} \right) -3.05(53), & r = 0.18\label{eq:aposrt} \\
    \log \left( \frac{R_P}{R_J} \right)  &= 0.02(5) \cdot {\rm[M/H]} - 0.79(1), & r = 0.02 \label{eq:aposrfe}  
\end{align}
in the ``small planet'' cluster (these clusters and the linear models are shown in Fig. \ref{fig:plan-apo}). Here, we can see that the APOGEE data reproduced the previously determined coefficients of   $T_{\rm eff}$ in both clusters, with larger ambiguities due to the significantly fewer planets in the currently availagble APOGEE sample. Also, the $r$ regression coefficients are smaller. In the case of the $R$--$T_{\rm eff}$ correlations, we see no significant dependence in the APOGEE data (the ambiguity of the coefficient of $T_{\rm eff}$ is larger than the value, hence, 0 is within the range) and the $r$ regression coefficients are also close to zero. Here, the APOGEE analysis reproduced the small correlation coefficients between $R_P$ and [M/H]. We note that as the uncertainties in the APOGEE data were also derived in a homogeneous way, we used these as weights for our regression.

The APOGEE data also contains the derived abundances\footnote{\url{https://www.sdss.org/dr17/irspec/abundances/}} for a number of elements. We also repeated the linear modeling described above for individual elements inculding C, C I (unionized C), N, O, Na, Mg, Al, Si, K, Ca, Mn, Co, Ni, Ce, and [C/O] to check for correlations between the relative abundance ratio of these elements and the planetary radius (Table \ref{tab:relative-apo}). In both clusters of ``larger'' and ``smaller'' planets, the relationship between the planetary radius and any given $X$ abundance is characterized by the slope of the line ($k$) and the intercept with the ordinate ($n$), such that:
\begin{equation}
    \log \frac{R_P}{R_J} = k \log \frac{X}{X_{\sun}} + n.
\end{equation}

To show an example of the fitted distributions, the left panel of Fig. \ref{fig:metacomp} shows the exoplanets in the $R_P$ -- $P$ plane with the metallicity as the coloring parameter taken from the APOGEE planet host sample. The right panel shows the residuals after removing the bilinear trend according to Eq. (\ref{eq:bilin}), which is very similar to the uncorrected one. The comparison of the two panels show that the significant features which can be partly explained by the metallicity includes a negative metallicity gradient along the period (at the boundary, stars with a metallicity higher than 0  are overabundant), mostly affecting the group of smaller planets, and a gradient along the radius. The large end of ``smaller planets'' and hot Jupiters tend to have increased metallicity. The appropriate KS-test here did not reveal differences of the metallicity distributions near and far from the edge of the desert. The plots for individual elements reproduced a very similar pattern, which we discuss below.

Table \ref{tab:relative-apo} shows those elements which we have found to be significantly correlated with the planetary radius within either the ``larger planet'' or the ``smaller planet'' sample. As a conclusion, O, Na, and K can influence the size distribution within the smaller planet group, while marginal significance was found in the case of N and Si in the ``smaller planet'' group as well. We found that only Mg influences the radius of planets in the ``large'' groups.

Correlations extending to the group of all planets, including both the "larger planet'' and the ``smaller planet'' samples are fitted with a bilinear regression in the period--radius plane. This equation is expressed as follows:
\begin{equation}\label{eq:bilin}
    \log \frac{X}{X_\sun} = a\cdot \log P + b \cdot \log \frac{R_P}{R_J} +c,
\end{equation}
where $a$, $b$, and $c$ are the coefficients of the linear regression. Table \ref{tab:bilin-apo} shows those elements where a significant ($p<0.05$) correlation was found. The most significant chemical parameter in the period--radius plane is the [M/H] itself, being very significant in both coordinates and we consider the overal metallicity as the most important forming parameter of the desert as well. The individual abundances, on the other hand, give us more insights to how the overall metallicity determines the chemical dependence of the desert boundaries.

The elements with significant effect of the period--radius distribution, besides what is explained by the [Fe/H], include C, N, O, Mg, Al, Si, K, and Mn. In addition, we did not find significant dependence involving Co and Ce (although these elements were determined only for 14 exoplanet-host stars), as well as [Ca/Fe], [Ni/Fe], and [C/O]. 

\begin{table*}
\caption{Fitted linear models to the distribution of exoplanets in the planetary radius-abundance parameter spaces for the two clusters.}
\label{tab:relative-apo}
\centering
\footnotesize
\begin{tabular}{l | c c c c c | c c c c c c c|}
\hline
\hline
Element & $N$\tablefootmark{1} & $k \pm \Delta k$ & $n \pm \Delta n$ & $r$\tablefootmark{2} & $p$\tablefootmark{3} & $N$\tablefootmark{1} & $k \pm \Delta k$ & $n \pm \Delta n$ & $r$\tablefootmark{2} & $p$\tablefootmark{3}  \\
\hline
& \multicolumn{5}{c}{\textit{Larger planets}} & \multicolumn{5}{c}{\textit{Smaller planets}} \\
\hline
% \text{[C/Fe]} & $106$ & $0.02 \pm 0.25$ & $-0.11 \pm -0.09$ & $0.03$ & $0.73$ & $568$ & $0.01 \pm 0.13$ & $-0.81 \pm -0.02$ & $0.01$ & $0.86$ \\ 
% \text{[C\textsc{i}/Fe]} & $106$ & $0.02 \pm 0.25$ & $-0.09 \pm -0.01$ & $0.01$ & $0.95$ & $567$ & $0.01 \pm 0.10$ & $-0.80 \pm -0.01$ & $0.00$ & $0.94$ \\ 
\text{[N/Fe]} & $106$ & $-0.04 \pm 0.11$ & $-0.10 \pm 0.03$ & $0.04$ & $0.72$ & $490$ & $0.11 \pm 0.06$ & $-0.79 \pm 0.01$ & $0.09$ & $0.06$ \\ 
\text{[O/Fe]} & $106$ & $0.35 \pm 0.23$ & $-0.13 \pm 0.02$ & $0.15$ & $0.12$ & $567$ & $0.21 \pm 0.10$ & $-0.83 \pm 0.01$ & $0.09$ & $0.03$ \\ 
\text{[Na/Fe]} & $85$ & $-0.08 \pm 0.08$ & $-0.77 \pm 0.09$ & $0.10$ & $0.34$ & $519$ & $0.20 \pm 0.08$ & $-0.67 \pm 0.02$ & $0.11$ & $0.01$ \\ 
\text{[Mg/Fe]} & $106$ & $-0.51 \pm 0.24$ & $-0.08 \pm 0.03$ & $0.20$ & $0.04$ & $568$ & $-0.03 \pm 0.09$ & $-0.79 \pm 0.01$ & $0.01$ & $0.77$ \\ 
 %\text{[Al/Fe]} & $101$ & $0.03 \pm 0.21$ & $-0.08 \pm 0.07$ & $0.03$ & $0.75$ & $514$ & $0.01 \pm 0.08$ & $-0.78 \pm -0.07$ & $0.04$ & $0.41$ \\ 
\text{[Si/Fe]} & $106$ & $0.30 \pm 0.32$ & $-0.10 \pm 0.03$ & $0.09$ & $0.35$ & $568$ & $0.22 \pm 0.12$ & $-0.80 \pm 0.01$ & $0.07$ & $0.08$ \\ 
\text{[K/Fe]} & $107$ & $-0.08 \pm 0.17$ & $-0.11 \pm 0.02$ & $0.04$ & $0.66$ & $566$ & $-0.16 \pm 0.07$ & $-0.80 \pm 0.01$ & $0.09$ & $0.03$ \\ 
 %\text{[Ca/Fe]} & $106$ & $0.02 \pm 0.33$ & $-0.09 \pm -0.27$ & $0.08$ & $0.41$ & $558$ & $0.01 \pm 0.14$ & $-0.79 \pm -0.09$ & $0.03$ & $0.51$ \\ 
 %\text{[Mn/Fe]} & $102$ & $0.02 \pm 0.21$ & $-0.07 \pm -0.01$ & $0.00$ & $0.98$ & $541$ & $0.01 \pm 0.13$ & $-0.79 \pm 0.16$ & $0.06$ & $0.20$ \\ 
 %\text{[Co/Fe]} & $4$ & $0.18 \pm 1.39$ & $-0.80 \pm 0.22$ & $0.11$ & $0.89$ & $10$ & $0.11 \pm 0.67$ & $-0.18 \pm 0.53$ & $0.27$ & $0.45$ \\ 
 %\text{[Ni/Fe]} & $106$ & $0.02 \pm 0.33$ & $-0.08 \pm 0.12$ & $0.03$ & $0.72$ & $568$ & $0.01 \pm 0.18$ & $-0.79 \pm 0.12$ & $0.03$ & $0.51$ \\ 
 %\text{[Ce/Fe]} & $5$ & $0.21 \pm 0.69$ & $-1.04 \pm -0.63$ & $0.47$ & $0.43$ & $9$ & $0.15 \pm 0.39$ & $0.03 \pm 0.49$ & $0.43$ & $0.24$ \\ 
% \text{[C/O]} & $106$ & $0.02 \pm 0.20$ & $-0.14 \pm -0.30$ & $0.15$ & $0.13$ & $567$ & $0.01 \pm 0.09$ & $-0.80 \pm -0.04$ & $0.02$ & $0.67$ \\ 
\hline
\end{tabular}
\tablefoot{
\tablefoottext{1}{Number of exoplanet hosts in the cluster.}
\tablefoottext{2}{Pearson's $r$ value.}
\tablefoottext{3}{$p$-values of the linear model.}
}
\end{table*}

A comparison of Tables \ref{tab:relative-apo} and \ref{tab:bilin-apo} shows that volatiles from the CNO process (especially N and O), moderately refractory alpha-process elements (Mg and Si) and K influence both the global structure of the planetary distribution -- surrounding the desert as well -- and they have positive correlations with the radius of smaller planets. Stars with increased N, O, Na, and Si, or decreased K abundances tend to form larger planets is the ``smaller planet'' groups. In the ``large planet'' group, only Mg is inversely correlated to the planet sizes. This revealed a set of elements that have a significant internal influence to the sizes of smaller and larger planets are disjointed, which serves as strong evidence that the formation and evolution of smaller and larger planets follow a chemical bimodality.

Table \ref{tab:bilin-apo} shows that the difference between the smaller/larger groups of planets and the global period--radius maps is related to mostly those elements that also appeared in Table \ref{tab:relative-apo}. Additional elements appear in Table \ref{tab:bilin-apo} as C, Al, and Mn. This also corroborates the interpretation that elements forming in CNO cycle or alpha process in stellar nucleosynthesis have an influence on the period-radius distribution of the planetary system. Interestingly, all refractory elements showed no or insignificant correlations in both tests, which elements (having the highest condensation temperatures, $T_{\rm cond}>1500$~K) are often considered as the initially condensed material in the protoplanetary disks \citep{2009GeCoA..73.5137S}, and the condensation center for less refractory and volatile materials in further steps. The distribution of planets in the period-radius plane and around the desert seems to be related to more or less volatile elements in the atmosphere, rather than the refractory elements condensed in the cores. Therefore, the desert itself appears to be a predominantly atmospheric feature, rather than a ``core'' feature, confirming the significance of atmospheric evaporation in its formation.

\begin{table}
\caption{Regression coefficients of Eq. (\ref{eq:bilin}).}
\label{tab:bilin-apo}
\centering
\scriptsize
\begin{tabular}{l c c c c c}
\hline
\hline
Element & $N$\tablefootmark{1} & $a \pm \Delta a$ & $b \pm \Delta b$ & $c \pm \Delta c$ & $p$\tablefootmark{2}  \\
\hline
\text{[M/H]} & $674$ & $-0.097 \pm 0.012$ & $0.198 \pm 0.022$ & $0.225 \pm 0.021$  & $< 10^{-4}$ \\ 
\text{[C/Fe]} & $674$ & $0.020 \pm 0.006$ & $-0.032 \pm 0.010$ & $-0.062 \pm 0.010$  & $0.0001$ \\ 
\text{[C I/Fe]} & $673$ & $0.027 \pm 0.006$ & $-0.040 \pm 0.011$ & $-0.082 \pm 0.011$  & $< 10^{-4}$ \\ 
\text{[N/Fe]} & $596$ & $0.010 \pm 0.015$ & $0.111 \pm 0.026$ & $0.116 \pm 0.026$  & $0.0001$ \\ 
\text{[O/Fe]} & $673$ & $0.023 \pm 0.008$ & $0.001 \pm 0.015$ & $0.040 \pm 0.015$  & $0.0164$ \\ 
%\text{[Na/Fe]} & $604$ & $-0.006 \pm 0.035$ & $0.019 \pm 0.062$ & $-0.057 \pm 0.061$  & $0.9454$ \\ 
\text{[Mg/Fe]} & $674$ & $0.030 \pm 0.006$ & $-0.031 \pm 0.011$ & $-0.030 \pm 0.011$  & $< 10^{-4}$ \\ 
 \text{[Al/Fe]}& $615$ & $0.021 \pm 0.008$ & $0.001 \pm 0.015$ & $0.095 \pm 0.015$  & $0.0430$ \\ 
\text{[Si/Fe]} & $674$ & $0.020 \pm 0.005$ & $0.016 \pm 0.009$ & $0.040 \pm 0.009$  & $< 10^{-4}$ \\ 
 \text{[K/Fe]}& $673$ & $0.026 \pm 0.009$ & $-0.035 \pm 0.016$ & $-0.022 \pm 0.015$  & $0.0025$ \\ 
 %\text{[Ca/Fe]}& $664$ & $0.008 \pm 0.005$ & $-0.012 \pm 0.009$ & $-0.021 \pm 0.008$  & $0.1382$ \\ 
 \text{[Mn/Fe]}& $643$ & $-0.019 \pm 0.006$ & $0.030 \pm 0.011$ & $0.049 \pm 0.011$  & $0.0005$ \\ 
%\text{[Co/Fe]} & $14$ & $-0.341 \pm 0.212$ & $-0.014 \pm 0.255$ & $0.402 \pm 0.250$  & $0.2957$ \\ 
% \text{[Ni/Fe]}& $674$ & $0.002 \pm 0.004$ & $0.004 \pm 0.007$ & $0.021 \pm 0.007$  & $0.6986$ \\ 
%\text{[Ce/Fe]} & $14$ & $0.184 \pm 0.153$ & $0.088 \pm 0.184$ & $-0.513 \pm 0.181$  & $0.4979$ \\ 
%\text{[C/O]} & $673$ & $-0.003 \pm 0.009$ & $-0.033 \pm 0.017$ & $-0.102 \pm 0.016$  & $0.1010$ \\ 
\hline
\end{tabular}
\tablefoot{
\tablefoottext{1}{Sample size.}
\tablefoottext{2}{$p$-value of the biliniear model.}
}
\end{table}

\section{Discussion}

Based on the data presented in Sects. \ref{sec:teff}--\ref{sec:meta}, we could generally confirm the results of \cite{2019MNRAS.485L.116S}. Most importantly, the dependence of the desert boundary on the stellar effective temperature, which had been interpreted as an evidence for the photoevaporation, have been confirmed in the present study as well. An important new detail to this early interpretation is the detection of a multiple and distinct population of hot Jupiters, having inflated and normal hot Jupiters at the boundary in the $R_P$--$P$ and the $M_P$--$P$ planes, respectively. Inflated hot Jupiters on the border can be a naturally seen as a new piece of evidence for the scenarios invoking photoevaporation \citep[e.g.,][]{2018MNRAS.479.5012O}, while the not inflated Jupiters at the boundary in the $R_P$--$P$ plane suggests a multiple track leading to the formation of the desert.

We also found that the previously claimed metallicity dependence of the border are in general less significant than expected previously. The earlier interpretation of the claimed detection was also interpreted in connection to the evaporation scenarios, as the more effective cooling of a metal-rich atmosphere was claimed to form more irradiation-endurant exoplanets. Our new results have demonstrated that planetary occurrence indeed reveals a dependence between the period, radius, and metallicity (in an agreement with the results of e.g. \cite{2018A&A...610A..63D} and \cite{petigura}), even though it has not been confirmed as a specific characteristic of the planets near the desert boundaries. 

The crowd of the exoplanets with spectroscopically detected atmospheric escape \citep{2021A&A...649A..40D} can also be considered as an evidence of irradiation effects forming the desert, because most of these planets are seen at the boundaries, or at least very close to it (Lecavelier des Etangs, personal communication). 
It can also be pointed out from Fig. \ref{fig:teff} that the hottest stars in our sample (with $T_{\rm eff} \sim 8000$ -- $10000$\;K) tend to host the largest and most massive hot Jupiters. A general trend that larger planets are more likely to be hosted by hotter stars can be observed from the median $T_{\rm eff}$ values from $A$, $B$, $C,$ and $D$ regions. Similar statements can be made upon examination of Figs. \ref{fig:rad} and \ref{fig:mass}; namely that stars with larger radius or mass are more likely to host larger planets. \cite{2021A&A...652A.110L} suggested that larger radii of planets surrounding larger stars are a result of different compositions: larger stars tend to host planets with larger H-He mass fractions. 
Stars with higher metallicities are likely to host larger planets, extending to periods as long as $\approx$20 days at least \citep{2018MNRAS.480.2206O}.
The explanation of the occurrence rate includes the cores of planets around more metal-rich stars to be more massive, hence, they can collect more initial atmospheric mass; and the photoevaporation of a more metal-rich atmosphere is more resistant to stellar irradiation due to the increased cooling related to the photoevaporation of such an atmosphere.

In \cite{2007A&A...461.1185L}, additional important factors have been introduced to the formalism, such as atmospheric loss due to tidal forces and the effect of he inclination. These two factors together have an influence on the lifetime of an evaporating atmosphere, leading to the conclusion that in the case of higher inclination values, increased density is required to retain the atmosphere during a specified lifetime. This result suggests that stars with higher temperatures (which tend to host more planets on high inclinations, \cite{2010ApJ...718L.145W,2012ApJ...757...18A,2015ARA&A..53..409W,2019AJ....158..141Z}) can host only gaseous planets with a higher average atmospheric density for a longer time. This conclusion would at least partly explain the dependence of the desert boundary on the stellar effective temperature, and the general dependence of planetary occurrence on metallicity. 

The most important result of our analysis is the detection of multiple relations between $M_P$ and $R_P$ on one side and $T_{\rm eff}$ and $M_S$ on the other side. These detections reflect the structural difference between large, atmosphere-dominated planets and the smaller Neptunes and super-Earths. These multiple relations complicate the picture at the desert boundary because it is precisely these two kinds of planets that meet there. 
There are different kind of planets at the upper boundary and the lower boundary, which can naturally lead to different mass and radius dependencies on the stellar parameters. However, in the case of the metallicity dependencies, the sign of the slope is also different and we see a negative correlation between $R_P$ and [M/H] for large planets, along with a positive correlation for small planets (Eqs. \ref{eq:grm} and \ref{eq:srm}). Therefore, we consider such multiple relations as an evidence for different processes forming the upper and lower boundaries of the desert (while there are at least two different processes at the upper boundary itself, as well; confirming the results from \cite{2018MNRAS.476.5639I}).

A widely claimed scenario here is the high-eccentricity migration \citep{2018MNRAS.479.5012O}, as well as the tidal disruption of giant planets \citep{2016ApJ...820L...8M} or an
XUV photoevaporation affecting sub-Saturns \citep{2022ApJ...924....9H}. Deciding which one is the dominant at the lower boundary of the desert requires detailed studies of individual planets. A possible detection of the ``sculpted Saturns'' scenario would corroborate the prediction of the possible surviving super low-density sub-Saturns to the present day if they are born with even larger atmospheres than they currently harbor -- in particular, Kepler 223 d has been claimed to be an example for such a planet on a 14.79~d period orbit \citep{2022ApJ...924....9H}. 

There are several other questions that we were not able to answer directly from the present data -- the most important being how the evaporation processes and the claimed high-eccentricity migration processes actually form the sub-Jovian/Neptunian desert. Also, because we found that the desert boundaries depend on stellar parameters, the simple linear and log-linear laws that mark the boundary of the desert \citep{2016A&A...589A..75M} are in some way transited to a multilinear dependence. Here, we find a lack of predictions about the desert boundary in the $R$-$P$ and $M$-$P$ planes as a function of stellar temperature and metallicity, for instance, and a comparison is not readily possible. Instead, our aim in the present analysis is to show that the formation of the desert is indeed a complex astrophysical process itself. Thus, presenting planetary evolution processes with respect to fundamental stellar parameters should indeed be the next step toward improving our understanding of this process.

Our current conclusions were based on the currently available NASA Exoplanet catalog, combined with the SDSS-IV APOGEE-2 catalog of exoplanet host stars with a limited number of entries ($<700$ planets in case of all elements). In the future, an extended stellar catalog from the SDSS-V Milky Way Mapper \citep[MWM, ][]{kollmeier_2017} program will cover a much larger sample of planet-host stars, and a similar analysis involving these data will reveal deeper details of these correlations.

\section{Summary}\label{sec:con}
In this paper, we analyzed exoplanet occurrence with respect to stellar and planetary parameters. Our  main results are as follows:

\begin{itemize}

\item{} We demonstrated the multifaced nature of the upper boundary of the sub-Jovian or Neptunian desert: in the $M_P$--$P$ plane, the boundary is marked by inflated hot Jupiters; while in the $R_P$--$P$ plane, there are normal hot Jupiters at the boundary.

\item{} We confirmed the dependence of the period boundary on stellar parameters, such as effective temperature, stellar mass, and stellar radius.

\item{} The multiple populations and the parameter dependence of the boundary suggests multiple formation mechanisms.

\item{} With fuzzy clustering, we also investigated double parameter relations between planet mass and radius on one side, and effective temperature, stellar mass, and metallicity on the other.

\item{} The demarcation of these planet groups coincides with the position of the desert, suggesting that all the relationships  connecting the planetary size and the fundamental stellar parameters  conjoin at the level of the size ranges of the desert and play a role in its formation.

\item{} In light of these results, we considered photoevaporation as the main process shaping the desert boundaries. We have discussed other possibilities such as high-eccentricity migration, abrasion by irradiation, tidal loss of atmosphere, and internal structural differences.

\end{itemize}

%\begin{figure}
 %   \centering
  %  \includegraphics[width = .48\textwidth]{HJ_shift.pdf}
   % \caption{Dual $y$ axis plot showing the sub-Jovian desert of exoplanets in both the $R_P - P$ and $M_P - P$ parameter space. The radii of planets are shown with black dots, while the masses are plotted with grey dots. Red and blue lines mark the shifting of the positions of exoplanets when switching from one plane to the other. The planets exactly on the border tend to change their position only slightly, while the hot Jupiters are considerably closer to the border of the desert in the $M_P - P$ plane.}
%    \label{fig:shift}
%\end{figure}

\begin{acknowledgements}
We acknowledge the support of the Hungarian National Research, Development and Innovation Office (NKFIH) grant K-125015, a PRODEX Experiment Agreement No. 4000137122, the Lend\"ulet LP2018-7/2022 grant of the Hungarian Academy of Science and the support of the city of Szombathely. LBo acknowledges the funding support from Italian Space Agency (ASI) regulated
by “Accordo ASI-INAF n. 2013-016-R.0 del 9 luglio 2013 e integrazione del 9 luglio 2015”. Prepared with the professional support of the Doctoral Student Scholarship Program of the Co-operative Doctoral Program of the Ministry of Innovation and Technology financed from the National Research, Development and Innovation Fund.
\end{acknowledgements}

\bibliographystyle{aa}
\bibliography{refs}

\begin{thebibliography}{64}
\expandafter\ifx\csname natexlab\endcsname\relax\def\natexlab#1{#1}\fi

\bibitem[{{Abdurro'uf} {et~al.}(2022){Abdurro'uf}, {Accetta}, {Aerts}, {Silva
  Aguirre}, {Ahumada}, {Ajgaonkar}, {Filiz Ak}, {Alam}, {Allende Prieto},
  {Almeida}, {Anders}, {Anderson}, {Andrews}, {Anguiano}, {Aquino-Ort{\'\i}z},
  {Arag{\'o}n-Salamanca}, {Argudo-Fern{\'a}ndez}, {Ata}, {Aubert},
  {Avila-Reese}, {Badenes}, {Barb{\'a}}, {Barger}, {Barrera-Ballesteros},
  {Beaton}, {Beers}, {Belfiore}, {Bender}, {Bernardi}, {Bershady}, {Beutler},
  {Bidin}, {Bird}, {Bizyaev}, {Blanc}, {Blanton}, {Boardman}, {Bolton},
  {Boquien}, {Borissova}, {Bovy}, {Brandt}, {Brown}, {Brownstein}, {Brusa},
  {Buchner}, {Bundy}, {Burchett}, {Bureau}, {Burgasser}, {Cabang}, {Campbell},
  {Cappellari}, {Carlberg}, {Wanderley}, {Carrera}, {Cash}, {Chen}, {Chen},
  {Cherinka}, {Chiappini}, {Choi}, {Chojnowski}, {Chung}, {Clerc}, {Cohen},
  {Comerford}, {Comparat}, {da Costa}, {Covey}, {Crane}, {Cruz-Gonzalez},
  {Culhane}, {Cunha}, {Dai}, {Damke}, {Darling}, {Davidson}, {Davies},
  {Dawson}, {De Lee}, {Diamond-Stanic}, {Cano-D{\'\i}az}, {S{\'a}nchez},
  {Donor}, {Duckworth}, {Dwelly}, {Eisenstein}, {Elsworth}, {Emsellem},
  {Eracleous}, {Escoffier}, {Fan}, {Farr}, {Feng}, {Fern{\'a}ndez-Trincado},
  {Feuillet}, {Filipp}, {Fillingham}, {Frinchaboy}, {Fromenteau}, {Galbany},
  {Garc{\'\i}a}, {Garc{\'\i}a-Hern{\'a}ndez}, {Ge}, {Geisler}, {Gelfand},
  {G{\'e}ron}, {Gibson}, {Goddy}, {Godoy-Rivera}, {Grabowski}, {Green},
  {Greener}, {Grier}, {Griffith}, {Guo}, {Guy}, {Hadjara}, {Harding},
  {Hasselquist}, {Hayes}, {Hearty}, {Hern{\'a}ndez}, {Hill}, {Hogg},
  {Holtzman}, {Horta}, {Hsieh}, {Hsu}, {Hsu}, {Huber}, {Huertas-Company},
  {Hutchinson}, {Hwang}, {Ibarra-Medel}, {Chitham}, {Ilha}, {Imig}, {Jaekle},
  {Jayasinghe}, {Ji}, {Johnson}, {Jones}, {J{\"o}nsson}, {Katkov}, {Khalatyan},
  {Kinemuchi}, {Kisku}, {Knapen}, {Kneib}, {Kollmeier}, {Kong}, {Kounkel},
  {Kreckel}, {Krishnarao}, {Lacerna}, {Lane}, {Langgin}, {Lavender}, {Law},
  {Lazarz}, {Leung}, {Leung}, {Lewis}, {Li}, {Li}, {Lian}, {Liang}, {Lin},
  {Lin}, {Lin}, {Lintott}, {Long}, {Longa-Pe{\~n}a}, {L{\'o}pez-Cob{\'a}},
  {Lu}, {Lundgren}, {Luo}, {Mackereth}, {de la Macorra}, {Mahadevan},
  {Majewski}, {Manchado}, {Mandeville}, {Maraston}, {Margalef-Bentabol},
  {Masseron}, {Masters}, {Mathur}, {McDermid}, {Mckay}, {Merloni},
  {Merrifield}, {Meszaros}, {Miglio}, {Di Mille}, {Minniti}, {Minsley},
  {Monachesi}, {Moon}, {Mosser}, {Mulchaey}, {Muna}, {Mu{\~n}oz}, {Myers},
  {Myers}, {Nadathur}, {Nair}, {Nandra}, {Neumann}, {Newman}, {Nidever},
  {Nikakhtar}, {Nitschelm}, {O'Connell}, {Garma-Oehmichen}, {Luan Souza de
  Oliveira}, {Olney}, {Oravetz}, {Ortigoza-Urdaneta}, {Osorio}, {Otter},
  {Pace}, {Padilla}, {Pan}, {Pan}, {Parikh}, {Parker}, {Peirani}, {Pe{\~n}a
  Ram{\'\i}rez}, {Penny}, {Percival}, {Perez-Fournon}, {Pinsonneault},
  {Poidevin}, {Poovelil}, {Price-Whelan}, {B{\'a}rbara de Andrade Queiroz},
  {Raddick}, {Ray}, {Rembold}, {Riddle}, {Riffel}, {Riffel}, {Rix}, {Robin},
  {Rodr{\'\i}guez-Puebla}, {Roman-Lopes}, {Rom{\'a}n-Z{\'u}{\~n}iga}, {Rose},
  {Ross}, {Rossi}, {Rubin}, {Salvato}, {S{\'a}nchez}, {S{\'a}nchez-Gallego},
  {Sanderson}, {Santana Rojas}, {Sarceno}, {Sarmiento}, {Sayres}, {Sazonova},
  {Schaefer}, {Schiavon}, {Schlegel}, {Schneider}, {Schultheis}, {Schwope},
  {Serenelli}, {Serna}, {Shao}, {Shapiro}, {Sharma}, {Shen}, {Shetrone}, {Shu},
  {Simon}, {Skrutskie}, {Smethurst}, {Smith}, {Sobeck}, {Spoo}, {Sprague},
  {Stark}, {Stassun}, {Steinmetz}, {Stello}, {Stone-Martinez},
  {Storchi-Bergmann}, {Stringfellow}, {Stutz}, {Su}, {Taghizadeh-Popp},
  {Talbot}, {Tayar}, {Telles}, {Teske}, {Thakar}, {Theissen}, {Tkachenko},
  {Thomas}, {Tojeiro}, {Hernandez Toledo}, {Troup}, {Trump}, {Trussler},
  {Turner}, {Tuttle}, {Unda-Sanzana}, {V{\'a}zquez-Mata}, {Valentini},
  {Valenzuela}, {Vargas-Gonz{\'a}lez}, {Vargas-Maga{\~n}a}, {Alfaro},
  {Villanova}, {Vincenzo}, {Wake}, {Warfield}, {Washington}, {Weaver},
  {Weijmans}, {Weinberg}, {Weiss}, {Westfall}, {Wild}, {Wilde}, {Wilson},
  {Wilson}, {Wilson}, {Wolf}, {Wood-Vasey}, {Yan}, {Zamora}, {Zasowski},
  {Zhang}, {Zhao}, {Zheng}, {Zheng}, \& {Zhu}}]{abdurro_2021}
{Abdurro'uf}, {Accetta}, K., {Aerts}, C., {et~al.} 2022, \apjs, 259, 35

\bibitem[{{Albrecht} {et~al.}(2012){Albrecht}, {Winn}, {Johnson}, {Howard},
  {Marcy}, {Butler}, {Arriagada}, {Crane}, {Shectman}, {Thompson}, {Hirano},
  {Bakos}, \& {Hartman}}]{2012ApJ...757...18A}
{Albrecht}, S., {Winn}, J.~N., {Johnson}, J.~A., {et~al.} 2012, \apj, 757, 18

\bibitem[{{Armstrong} {et~al.}(2020){Armstrong}, {Lopez}, {Adibekyan}, {Booth},
  {Bryant}, {Collins}, {Deleuil}, {Emsenhuber}, {Huang}, {King}, {Lillo-Box},
  {Lissauer}, {Matthews}, {Mousis}, {Nielsen}, {Osborn}, {Otegi}, {Santos},
  {Sousa}, {Stassun}, {Veras}, {Ziegler}, {Acton}, {Almenara}, {Anderson},
  {Barrado}, {Barros}, {Bayliss}, {Belardi}, {Bouchy}, {Brice{\~n}o}, {Brogi},
  {Brown}, {Burleigh}, {Casewell}, {Chaushev}, {Ciardi}, {Collins},
  {Col{\'o}n}, {Cooke}, {Crossfield}, {D{\'\i}az}, {Delgado Mena}, {Demangeon},
  {Dorn}, {Dumusque}, {Eigm{\"u}ller}, {Fausnaugh}, {Figueira}, {Gan},
  {Gandhi}, {Gill}, {Gonzales}, {Goad}, {G{\"u}nther}, {Helled}, {Hojjatpanah},
  {Howell}, {Jackman}, {Jenkins}, {Jenkins}, {Jensen}, {Kennedy}, {Latham},
  {Law}, {Lendl}, {Lozovsky}, {Mann}, {Moyano}, {McCormac}, {Meru},
  {Mordasini}, {Osborn}, {Pollacco}, {Queloz}, {Raynard}, {Ricker}, {Rowden},
  {Santerne}, {Schlieder}, {Seager}, {Sha}, {Tan}, {Tilbrook}, {Ting}, {Udry},
  {Vanderspek}, {Watson}, {West}, {Wilson}, {Winn}, {Wheatley}, {Villasenor},
  {Vines}, \& {Zhan}}]{2020Natur.583...39A}
{Armstrong}, D.~J., {Lopez}, T.~A., {Adibekyan}, V., {et~al.} 2020, \nat, 583,
  39

\bibitem[{{Ataiee} \& {Kley}(2021)}]{2021A&A...648A..69A}
{Ataiee}, S. \& {Kley}, W. 2021, \aap, 648, A69

\bibitem[{{Bailey} \& {Batygin}(2018)}]{2018ApJ...866L...2B}
{Bailey}, E. \& {Batygin}, K. 2018, \apjl, 866, L2

\bibitem[{{Ben{\'\i}tez-Llambay} {et~al.}(2011){Ben{\'\i}tez-Llambay},
  {Masset}, \& {Beaug{\'e}}}]{2011A&A...528A...2B}
{Ben{\'\i}tez-Llambay}, P., {Masset}, F., \& {Beaug{\'e}}, C. 2011, \aap, 528,
  A2

\bibitem[{Bezdek(1981)}]{bezdek}
Bezdek, J.~C. 1981, Pattern recognition with fuzzy objective function
  algorithms. (Springer New York, NY)

\bibitem[{{Blanton} {et~al.}(2017){Blanton}, {Bershady}, {Abolfathi},
  {Albareti}, {Allende Prieto}, {Almeida}, {Alonso-Garc{\'\i}a}, {Anders},
  {Anderson}, {Andrews}, {Aquino-Ort{\'\i}z}, {Arag{\'o}n-Salamanca},
  {Argudo-Fern{\'a}ndez}, {Armengaud}, {Aubourg}, {Avila-Reese}, {Badenes},
  {Bailey}, {Barger}, {Barrera-Ballesteros}, {Bartosz}, {Bates}, {Baumgarten},
  {Bautista}, {Beaton}, {Beers}, {Belfiore}, {Bender}, {Berlind}, {Bernardi},
  {Beutler}, {Bird}, {Bizyaev}, {Blanc}, {Blomqvist}, {Bolton}, {Boquien},
  {Borissova}, {van den Bosch}, {Bovy}, {Brandt}, {Brinkmann}, {Brownstein},
  {Bundy}, {Burgasser}, {Burtin}, {Busca}, {Cappellari}, {Delgado Carigi},
  {Carlberg}, {Carnero Rosell}, {Carrera}, {Chanover}, {Cherinka}, {Cheung},
  {G{\'o}mez Maqueo Chew}, {Chiappini}, {Choi}, {Chojnowski}, {Chuang},
  {Chung}, {Cirolini}, {Clerc}, {Cohen}, {Comparat}, {da Costa}, {Cousinou},
  {Covey}, {Crane}, {Croft}, {Cruz-Gonzalez}, {Garrido Cuadra}, {Cunha},
  {Damke}, {Darling}, {Davies}, {Dawson}, {de la Macorra}, {Dell'Agli}, {De
  Lee}, {Delubac}, {Di Mille}, {Diamond-Stanic}, {Cano-D{\'\i}az}, {Donor},
  {Downes}, {Drory}, {du Mas des Bourboux}, {Duckworth}, {Dwelly}, {Dyer},
  {Ebelke}, {Eigenbrot}, {Eisenstein}, {Emsellem}, {Eracleous}, {Escoffier},
  {Evans}, {Fan}, {Fern{\'a}ndez-Alvar}, {Fernandez-Trincado}, {Feuillet},
  {Finoguenov}, {Fleming}, {Font-Ribera}, {Fredrickson}, {Freischlad},
  {Frinchaboy}, {Fuentes}, {Galbany}, {Garcia-Dias},
  {Garc{\'\i}a-Hern{\'a}ndez}, {Gaulme}, {Geisler}, {Gelfand},
  {Gil-Mar{\'\i}n}, {Gillespie}, {Goddard}, {Gonzalez-Perez}, {Grabowski},
  {Green}, {Grier}, {Gunn}, {Guo}, {Guy}, {Hagen}, {Hahn}, {Hall}, {Harding},
  {Hasselquist}, {Hawley}, {Hearty}, {Gonzalez Hern{\'a}ndez}, {Ho}, {Hogg},
  {Holley-Bockelmann}, {Holtzman}, {Holzer}, {Huehnerhoff}, {Hutchinson},
  {Hwang}, {Ibarra-Medel}, {da Silva Ilha}, {Ivans}, {Ivory}, {Jackson},
  {Jensen}, {Johnson}, {Jones}, {J{\"o}nsson}, {Jullo}, {Kamble}, {Kinemuchi},
  {Kirkby}, {Kitaura}, {Klaene}, {Knapp}, {Kneib}, {Kollmeier}, {Lacerna},
  {Lane}, {Lang}, {Law}, {Lazarz}, {Lee}, {Le Goff}, {Liang}, {Li}, {Li},
  {Lian}, {Lima}, {Lin}, {Lin}, {Bertran de Lis}, {Liu}, {de Icaza Lizaola},
  {Long}, {Lucatello}, {Lundgren}, {MacDonald}, {Deconto Machado}, {MacLeod},
  {Mahadevan}, {Geimba Maia}, {Maiolino}, {Majewski}, {Malanushenko},
  {Malanushenko}, {Manchado}, {Mao}, {Maraston}, {Marques-Chaves}, {Masseron},
  {Masters}, {McBride}, {McDermid}, {McGrath}, {McGreer}, {Medina Pe{\~n}a},
  {Melendez}, {Merloni}, {Merrifield}, {Meszaros}, {Meza}, {Minchev},
  {Minniti}, {Miyaji}, {More}, {Mulchaey}, {M{\"u}ller-S{\'a}nchez}, {Muna},
  {Munoz}, {Myers}, {Nair}, {Nandra}, {Correa do Nascimento}, {Negrete},
  {Ness}, {Newman}, {Nichol}, {Nidever}, {Nitschelm}, {Ntelis}, {O'Connell},
  {Oelkers}, {Oravetz}, {Oravetz}, {Pace}, {Padilla}, {Palanque-Delabrouille},
  {Alonso Palicio}, {Pan}, {Parejko}, {Parikh}, {P{\^a}ris}, {Park}, {Patten},
  {Peirani}, {Pellejero-Ibanez}, {Penny}, {Percival}, {Perez-Fournon},
  {Petitjean}, {Pieri}, {Pinsonneault}, {Pisani}, {Poleski}, {Prada},
  {Prakash}, {Queiroz}, {Raddick}, {Raichoor}, {Barboza Rembold}, {Richstein},
  {Riffel}, {Riffel}, {Rix}, {Robin}, {Rockosi}, {Rodr{\'\i}guez-Torres},
  {Roman-Lopes}, {Rom{\'a}n-Z{\'u}{\~n}iga}, {Rosado}, {Ross}, {Rossi}, {Ruan},
  {Ruggeri}, {Rykoff}, {Salazar-Albornoz}, {Salvato}, {S{\'a}nchez}, {Aguado},
  {S{\'a}nchez-Gallego}, {Santana}, {Santiago}, {Sayres}, {Schiavon}, {da Silva
  Schimoia}, {Schlafly}, {Schlegel}, {Schneider}, {Schultheis}, {Schuster},
  {Schwope}, {Seo}, {Shao}, {Shen}, {Shetrone}, {Shull}, {Simon}, {Skinner},
  {Skrutskie}, {Slosar}, {Smith}, {Sobeck}, {Sobreira}, {Somers}, {Souto},
  {Stark}, {Stassun}, {Stauffer}, {Steinmetz}, {Storchi-Bergmann},
  {Streblyanska}, {Stringfellow}, {Su{\'a}rez}, {Sun}, {Suzuki}, {Szigeti},
  {Taghizadeh-Popp}, {Tang}, {Tao}, {Tayar}, {Tembe}, {Teske}, {Thakar},
  {Thomas}, {Thompson}, {Tinker}, {Tissera}, {Tojeiro}, {Hernandez Toledo}, {de
  la Torre}, {Tremonti}, {Troup}, {Valenzuela}, {Martinez Valpuesta},
  {Vargas-Gonz{\'a}lez}, {Vargas-Maga{\~n}a}, {Vazquez}, {Villanova}, {Vivek},
  {Vogt}, {Wake}, {Walterbos}, {Wang}, {Weaver}, {Weijmans}, {Weinberg},
  {Westfall}, {Whelan}, {Wild}, {Wilson}, {Wood-Vasey}, {Wylezalek}, {Xiao},
  {Yan}, {Yang}, {Ybarra}, {Y{\`e}che}, {Zakamska}, {Zamora}, {Zarrouk},
  {Zasowski}, {Zhang}, {Zhao}, {Zheng}, {Zheng}, {Zhou}, {Zhou}, {Zhu},
  {Zoccali}, \& {Zou}}]{blanton_2017}
{Blanton}, M.~R., {Bershady}, M.~A., {Abolfathi}, B., {et~al.} 2017, \aj, 154,
  28

\bibitem[{{Col{\'o}n} {et~al.}(2015){Col{\'o}n}, {Morehead}, \&
  {Ford}}]{2015MNRAS.452.3001C}
{Col{\'o}n}, K.~D., {Morehead}, R.~C., \& {Ford}, E.~B. 2015, \mnras, 452, 3001

\bibitem[{{Demangeon} {et~al.}(2018){Demangeon}, {Faedi}, {H{\'e}brard},
  {Brown}, {Barros}, {Doyle}, {Maxted}, {Collier Cameron}, {Hay}, {Alikakos},
  {Anderson}, {Armstrong}, {Boumis}, {Bonomo}, {Bouchy}, {Delrez}, {Gillon},
  {Haswell}, {Hellier}, {Jehin}, {Kiefer}, {Lam}, {Lendl}, {Mancini},
  {McCormac}, {Norton}, {Osborn}, {Palle}, {Pepe}, {Pollacco}, {Prieto-Arranz},
  {Queloz}, {S{\'e}gransan}, {Smalley}, {Triaud}, {Udry}, {West}, \&
  {Wheatley}}]{2018A&A...610A..63D}
{Demangeon}, O.~D.~S., {Faedi}, F., {H{\'e}brard}, G., {et~al.} 2018, \aap,
  610, A63

\bibitem[{{Dong} {et~al.}(2018){Dong}, {Xie}, {Zhou}, {Zheng}, \& {Luo}}]{dong}
{Dong}, S., {Xie}, J.-W., {Zhou}, J.-L., {Zheng}, Z., \& {Luo}, A. 2018,
  Proceedings of the National Academy of Science, 115, 266

\bibitem[{{dos Santos} {et~al.}(2021){dos Santos}, {Bourrier}, {Ehrenreich},
  {Sanz-Forcada}, {L{\'o}pez-Morales}, {Sing}, {Garc{\'\i}a Mu{\~n}oz},
  {Henry}, {Lavvas}, {Lecavelier des Etangs}, {Mikal-Evans}, {Vidal-Madjar}, \&
  {Wakeford}}]{2021A&A...649A..40D}
{dos Santos}, L.~A., {Bourrier}, V., {Ehrenreich}, D., {et~al.} 2021, \aap,
  649, A40

\bibitem[{{Dragomir} {et~al.}(2020){Dragomir}, {Crossfield}, {Benneke}, {Wong},
  {Daylan}, {Diaz}, {Deming}, {Molliere}, {Kreidberg}, {Jenkins}, {Berardo},
  {Christiansen}, {Dressing}, {Gorjian}, {Kane}, {Mikal-Evans}, {Morales},
  {Werner}, {Ricker}, {Vanderspek}, {Seager}, {Winn}, {Jenkins}, {Col{\'o}n},
  {Fong}, {Guerrero}, {Hesse}, {Osborn}, {E. Rose}, {Smith}, \&
  {Ting}}]{2020ApJ...903L...6D}
{Dragomir}, D., {Crossfield}, I. J.~M., {Benneke}, B., {et~al.} 2020, \apjl,
  903, L6

\bibitem[{{Eigm{\"u}ller} {et~al.}(2017){Eigm{\"u}ller}, {Gandolfi}, {Persson},
  {Donati}, {Fridlund}, {Csizmadia}, {Barrag{\'a}n}, {Smith}, {Cabrera},
  {Korth}, {Grziwa}, {Prieto-Arranz}, {Nespral}, {Saario}, {Cochran}, {Cusano},
  {Deeg}, {Endl}, {Erikson}, {Guenther}, {Hatzes}, {P{\"a}tzold}, \&
  {Rauer}}]{2017AJ....153..130E}
{Eigm{\"u}ller}, P., {Gandolfi}, D., {Persson}, C.~M., {et~al.} 2017, \aj, 153,
  130

\bibitem[{{Emsenhuber} {et~al.}(2021{\natexlab{a}}){Emsenhuber}, {Mordasini},
  {Burn}, {Alibert}, {Benz}, \& {Asphaug}}]{2021A&A...656A..69E}
{Emsenhuber}, A., {Mordasini}, C., {Burn}, R., {et~al.} 2021{\natexlab{a}},
  \aap, 656, A69

\bibitem[{{Emsenhuber} {et~al.}(2021{\natexlab{b}}){Emsenhuber}, {Mordasini},
  {Burn}, {Alibert}, {Benz}, \& {Asphaug}}]{2021A&A...656A..70E}
{Emsenhuber}, A., {Mordasini}, C., {Burn}, R., {et~al.} 2021{\natexlab{b}},
  \aap, 656, A70

\bibitem[{Feigelson \& Babu(2012)}]{feigelson_babu_2012}
Feigelson, E.~D. \& Babu, G.~J. 2012, Modern Statistical Methods for Astronomy:
  With R Applications (Cambridge University Press)

\bibitem[{Ferraro {et~al.}(2019)Ferraro, Giordani, \& Serafini}]{fclust}
Ferraro, M., Giordani, P., \& Serafini, A. 2019, The {R} Journal, 11

\bibitem[{{Fortney} {et~al.}(2007){Fortney}, {Marley}, \& {Barnes}}]{fortney}
{Fortney}, J.~J., {Marley}, M.~S., \& {Barnes}, J.~W. 2007, \apj, 659, 1661

\bibitem[{{Giacalone} {et~al.}(2017){Giacalone}, {Matsakos}, \&
  {K{\"o}nigl}}]{2017AJ....154..192G}
{Giacalone}, S., {Matsakos}, T., \& {K{\"o}nigl}, A. 2017, \aj, 154, 192

\bibitem[{{Gunn} {et~al.}(2006){Gunn}, {Siegmund}, {Mannery}, {Owen}, {Hull},
  {Leger}, {Carey}, {Knapp}, {York}, {Boroski}, {Kent}, {Lupton}, {Rockosi},
  {Evans}, {Waddell}, {Anderson}, {Annis}, {Barentine}, {Bartoszek}, {Bastian},
  {Bracker}, {Brewington}, {Briegel}, {Brinkmann}, {Brown}, {Carr},
  {Czarapata}, {Drennan}, {Dombeck}, {Federwitz}, {Gillespie}, {Gonzales},
  {Hansen}, {Harvanek}, {Hayes}, {Jordan}, {Kinney}, {Klaene}, {Kleinman},
  {Kron}, {Kresinski}, {Lee}, {Limmongkol}, {Lindenmeyer}, {Long}, {Loomis},
  {McGehee}, {Mantsch}, {Neilsen}, {Neswold}, {Newman}, {Nitta}, {Peoples},
  {Pier}, {Prieto}, {Prosapio}, {Rivetta}, {Schneider}, {Snedden}, \&
  {Wang}}]{gunn_2006}
{Gunn}, J.~E., {Siegmund}, W.~A., {Mannery}, E.~J., {et~al.} 2006, \aj, 131,
  2332

\bibitem[{{Hallatt} \& {Lee}(2022)}]{2022ApJ...924....9H}
{Hallatt}, T. \& {Lee}, E.~J. 2022, \apj, 924, 9

\bibitem[{{Ionov} {et~al.}(2018){Ionov}, {Pavlyuchenkov}, \&
  {Shematovich}}]{2018MNRAS.476.5639I}
{Ionov}, D.~E., {Pavlyuchenkov}, Y.~N., \& {Shematovich}, V.~I. 2018, \mnras,
  476, 5639

\bibitem[{{Jenkins} {et~al.}(2020){Jenkins}, {D{\'\i}az}, {Kurtovic},
  {Espinoza}, {Vines}, {Rojas}, {Brahm}, {Torres}, {Cort{\'e}s-Zuleta}, {Soto},
  {Lopez}, {King}, {Wheatley}, {Winn}, {Ciardi}, {Ricker}, {Vanderspek},
  {Latham}, {Seager}, {Jenkins}, {Beichman}, {Bieryla}, {Burke},
  {Christiansen}, {Henze}, {Klaus}, {McCauliff}, {Mori}, {Narita}, {Nishiumi},
  {Tamura}, {de Leon}, {Quinn}, {Villase{\~n}or}, {Vezie}, {Lissauer},
  {Collins}, {Collins}, {Isopi}, {Mallia}, {Ercolino}, {Petrovich},
  {Jord{\'a}n}, {Acton}, {Armstrong}, {Bayliss}, {Bouchy}, {Belardi}, {Bryant},
  {Burleigh}, {Cabrera}, {Casewell}, {Chaushev}, {Cooke}, {Eigm{\"u}ller},
  {Erikson}, {Foxell}, {G{\"a}nsicke}, {Gill}, {Gillen}, {G{\"u}nther}, {Goad},
  {Hooton}, {Jackman}, {Louden}, {McCormac}, {Moyano}, {Nielsen}, {Pollacco},
  {Queloz}, {Rauer}, {Raynard}, {Smith}, {Tilbrook}, {Titz-Weider}, {Turner},
  {Udry}, {Walker}, {Watson}, {West}, {Palle}, {Ziegler}, {Law}, \&
  {Mann}}]{2020NatAs...4.1148J}
{Jenkins}, J.~S., {D{\'\i}az}, M.~R., {Kurtovic}, N.~T., {et~al.} 2020, Nature
  Astronomy, 4, 1148

\bibitem[{{Kollmeier} {et~al.}(2017){Kollmeier}, {Zasowski}, {Rix}, {Johns},
  {Anderson}, {Drory}, {Johnson}, {Pogge}, {Bird}, {Blanc}, {Brownstein},
  {Crane}, {De Lee}, {Klaene}, {Kreckel}, {MacDonald}, {Merloni}, {Ness},
  {O'Brien}, {Sanchez-Gallego}, {Sayres}, {Shen}, {Thakar}, {Tkachenko},
  {Aerts}, {Blanton}, {Eisenstein}, {Holtzman}, {Maoz}, {Nandra}, {Rockosi},
  {Weinberg}, {Bovy}, {Casey}, {Chaname}, {Clerc}, {Conroy}, {Eracleous},
  {G{\"a}nsicke}, {Hekker}, {Horne}, {Kauffmann}, {McQuinn}, {Pellegrini},
  {Schinnerer}, {Schlafly}, {Schwope}, {Seibert}, {Teske}, \& {van
  Saders}}]{kollmeier_2017}
{Kollmeier}, J.~A., {Zasowski}, G., {Rix}, H.-W., {et~al.} 2017, arXiv
  e-prints, arXiv:1711.03234

\bibitem[{{Kurokawa} \& {Nakamoto}(2014)}]{2014ApJ...783...54K}
{Kurokawa}, H. \& {Nakamoto}, T. 2014, \apj, 783, 54

\bibitem[{{Lecavelier Des Etangs}(2007)}]{2007A&A...461.1185L}
{Lecavelier Des Etangs}, A. 2007, \aap, 461, 1185

\bibitem[{{Lin} \& {Papaloizou}(1986)}]{1986ApJ...309..846L}
{Lin}, D.~N.~C. \& {Papaloizou}, J. 1986, \apj, 309, 846

\bibitem[{{Lopez} \& {Fortney}(2014)}]{2014ApJ...792....1L}
{Lopez}, E.~D. \& {Fortney}, J.~J. 2014, \apj, 792, 1

\bibitem[{{Lozovsky} {et~al.}(2021){Lozovsky}, {Helled}, {Pascucci}, {Dorn},
  {Venturini}, \& {Feldmann}}]{2021A&A...652A.110L}
{Lozovsky}, M., {Helled}, R., {Pascucci}, I., {et~al.} 2021, \aap, 652, A110

\bibitem[{{Lundkvist} {et~al.}(2016){Lundkvist}, {Kjeldsen}, {Albrecht},
  {Davies}, {Basu}, {Huber}, {Justesen}, {Karoff}, {Silva Aguirre}, {van
  Eylen}, {Vang}, {Arentoft}, {Barclay}, {Bedding}, {Campante}, {Chaplin},
  {Christensen-Dalsgaard}, {Elsworth}, {Gilliland}, {Handberg}, {Hekker},
  {Kawaler}, {Lund}, {Metcalfe}, {Miglio}, {Rowe}, {Stello}, {Tingley}, \&
  {White}}]{2016NatCo...711201L}
{Lundkvist}, M.~S., {Kjeldsen}, H., {Albrecht}, S., {et~al.} 2016, Nature
  Communications, 7, 11201

\bibitem[{{Majewski} {et~al.}(2017){Majewski}, {Schiavon}, {Frinchaboy},
  {Allende Prieto}, {Barkhouser}, {Bizyaev}, {Blank}, {Brunner}, {Burton},
  {Carrera}, {Chojnowski}, {Cunha}, {Epstein}, {Fitzgerald}, {Garc{\'\i}a
  P{\'e}rez}, {Hearty}, {Henderson}, {Holtzman}, {Johnson}, {Lam}, {Lawler},
  {Maseman}, {M{\'e}sz{\'a}ros}, {Nelson}, {Nguyen}, {Nidever}, {Pinsonneault},
  {Shetrone}, {Smee}, {Smith}, {Stolberg}, {Skrutskie}, {Walker}, {Wilson},
  {Zasowski}, {Anders}, {Basu}, {Beland}, {Blanton}, {Bovy}, {Brownstein},
  {Carlberg}, {Chaplin}, {Chiappini}, {Eisenstein}, {Elsworth}, {Feuillet},
  {Fleming}, {Galbraith-Frew}, {Garc{\'\i}a}, {Garc{\'\i}a-Hern{\'a}ndez},
  {Gillespie}, {Girardi}, {Gunn}, {Hasselquist}, {Hayden}, {Hekker}, {Ivans},
  {Kinemuchi}, {Klaene}, {Mahadevan}, {Mathur}, {Mosser}, {Muna}, {Munn},
  {Nichol}, {O'Connell}, {Parejko}, {Robin}, {Rocha-Pinto}, {Schultheis},
  {Serenelli}, {Shane}, {Silva Aguirre}, {Sobeck}, {Thompson}, {Troup},
  {Weinberg}, \& {Zamora}}]{majewski_2017}
{Majewski}, S.~R., {Schiavon}, R.~P., {Frinchaboy}, P.~M., {et~al.} 2017, \aj,
  154, 94

\bibitem[{{Masset} {et~al.}(2006){Masset}, {Morbidelli}, {Crida}, \&
  {Ferreira}}]{2006ApJ...642..478M}
{Masset}, F.~S., {Morbidelli}, A., {Crida}, A., \& {Ferreira}, J. 2006, \apj,
  642, 478

\bibitem[{{Matsakos} \& {K{\"o}nigl}(2016)}]{2016ApJ...820L...8M}
{Matsakos}, T. \& {K{\"o}nigl}, A. 2016, \apjl, 820, L8

\bibitem[{{Mazeh} {et~al.}(2016){Mazeh}, {Holczer}, \&
  {Faigler}}]{2016A&A...589A..75M}
{Mazeh}, T., {Holczer}, T., \& {Faigler}, S. 2016, \aap, 589, A75

\bibitem[{{Mordasini} {et~al.}(2012){Mordasini}, {Alibert}, {Georgy},
  {Dittkrist}, {Klahr}, \& {Henning}}]{2012A&A...547A.112M}
{Mordasini}, C., {Alibert}, Y., {Georgy}, C., {et~al.} 2012, \aap, 547, A112

\bibitem[{{Mordasini} {et~al.}(2015){Mordasini}, {Molli{\`e}re}, {Dittkrist},
  {Jin}, \& {Alibert}}]{2015IJAsB..14..201M}
{Mordasini}, C., {Molli{\`e}re}, P., {Dittkrist}, K.~M., {Jin}, S., \&
  {Alibert}, Y. 2015, International Journal of Astrobiology, 14, 201

\bibitem[{{Mori} {et~al.}(2022){Mori}, {Livingston}, {Leon}, {Narita},
  {Hirano}, {Fukui}, {Collins}, {Fujita}, {Hori}, {Ishikawa}, {Kawauchi},
  {Stassun}, {Watanabe}, {Giacalone}, {Gore}, {Schroeder}, {Dressing},
  {Bieryla}, {Jensen}, {Massey}, {Shporer}, {Kuzuhara}, {Charbonneau},
  {Ciardi}, {Doty}, {Esparza-Borges}, {Harakawa}, {Hodapp}, {Ikoma}, {Ikuta},
  {Isogai}, {Jenkins}, {Kagetani}, {Kimura}, {Kodama}, {Kotani},
  {Krishnamurthy}, {Kudo}, {Kurita}, {Kurokawa}, {Kusakabe}, {Latham},
  {McLean}, {Murgas}, {Nishikawa}, {Nishiumi}, {Omiya}, {Osborn}, {Palle},
  {Parviainen}, {Ricker}, {Seager}, {Serizawa}, {Teng}, {Terada}, {Twicken},
  {Ueda}, {Vanderspek}, {Vievard}, {Winn}, {Zou}, \&
  {Tamura}}]{2022AJ....163..298M}
{Mori}, M., {Livingston}, J.~H., {Leon}, J.~d., {et~al.} 2022, \aj, 163, 298

\bibitem[{{Murgas} {et~al.}(2021){Murgas}, {Astudillo-Defru}, {Bonfils},
  {Crossfield}, {Almenara}, {Livingston}, {Stassun}, {Korth}, {Orell-Miquel},
  {Morello}, {Eastman}, {Lissauer}, {Kane}, {Morales}, {Werner}, {Gorjian},
  {Benneke}, {Dragomir}, {Matthews}, {Howell}, {Ciardi}, {Gonzales}, {Matson},
  {Beichman}, {Schlieder}, {Collins}, {Collins}, {Jensen}, {Evans}, {Pozuelos},
  {Gillon}, {Jehin}, {Barkaoui}, {Artigau}, {Bouchy}, {Charbonneau},
  {Delfosse}, {D{\'\i}az}, {Doyon}, {Figueira}, {Forveille}, {Lovis}, {Melo},
  {Gaisn{\'e}}, {Pepe}, {Santos}, {S{\'e}gransan}, {Udry}, {Goeke}, {Levine},
  {Quintana}, {Guerrero}, {Mireles}, {Caldwell}, {Tenenbaum}, {Brasseur},
  {Ricker}, {Vanderspek}, {Latham}, {Seager}, {Winn}, \&
  {Jenkins}}]{2021A&A...653A..60M}
{Murgas}, F., {Astudillo-Defru}, N., {Bonfils}, X., {et~al.} 2021, \aap, 653,
  A60

\bibitem[{{Owen}(2019)}]{2019AREPS..47...67O}
{Owen}, J.~E. 2019, Annual Review of Earth and Planetary Sciences, 47, 67

\bibitem[{{Owen} \& {Lai}(2018)}]{2018MNRAS.479.5012O}
{Owen}, J.~E. \& {Lai}, D. 2018, \mnras, 479, 5012

\bibitem[{{Owen} \& {Murray-Clay}(2018)}]{2018MNRAS.480.2206O}
{Owen}, J.~E. \& {Murray-Clay}, R. 2018, \mnras, 480, 2206

\bibitem[{{Owen} \& {Wu}(2016)}]{2016ApJ...817..107O}
{Owen}, J.~E. \& {Wu}, Y. 2016, \apj, 817, 107

\bibitem[{{Persson} {et~al.}(2022){Persson}, {Georgieva}, {Gandolfi}, {Acuna},
  {Aguichine}, {Muresan}, {Guenther}, {Livingston}, {Collins}, {Dai},
  {Fridlund}, {Goffo}, {Jenkins}, {Kab{\'a}th}, {Korth}, {Levine}, {Serrano},
  {Vines}, {Barragan}, {Carleo}, {Colon}, {Cochran}, {Christiansen}, {Deeg},
  {Deleuil}, {Dragomir}, {Esposito}, {Gan}, {Grziwa}, {Hatzes}, {Hesse},
  {Horne}, {Jenkins}, {Kielkopf}, {Klagyivik}, {Lam}, {Latham}, {Luque},
  {Orell-Miquel}, {Mortier}, {Mousis}, {Narita}, {Osborne}, {Palle}, {Papini},
  {Ricker}, {Schmerling}, {Seager}, {Stassun}, {Van Eylen}, {Vanderspek},
  {Wang}, {Winn}, {Wohler}, {Zambelli}, \& {Ziegler}}]{2022A&A...666A.184P}
{Persson}, C.~M., {Georgieva}, I.~Y., {Gandolfi}, D., {et~al.} 2022, \aap, 666,
  A184

\bibitem[{{Petigura} {et~al.}(2018){Petigura}, {Marcy}, {Winn}, {Weiss},
  {Fulton}, {Howard}, {Sinukoff}, {Isaacson}, {Morton}, \&
  {Johnson}}]{petigura}
{Petigura}, E.~A., {Marcy}, G.~W., {Winn}, J.~N., {et~al.} 2018, \aj, 155, 89

\bibitem[{{Rauer} {et~al.}(2014){Rauer}, {Catala}, {Aerts}, {Appourchaux},
  {Benz}, {Brandeker}, {Christensen-Dalsgaard}, {Deleuil}, {Gizon}, {Goupil},
  {G{\"u}del}, {Janot-Pacheco}, {Mas-Hesse}, {Pagano}, {Piotto}, {Pollacco},
  {Santos}, {Smith}, {Su{\'a}rez}, {Szab{\'o}}, {Udry}, {Adibekyan}, {Alibert},
  {Almenara}, {Amaro-Seoane}, {Eiff}, {Asplund}, {Antonello}, {Barnes},
  {Baudin}, {Belkacem}, {Bergemann}, {Bihain}, {Birch}, {Bonfils}, {Boisse},
  {Bonomo}, {Borsa}, {Brand{\~a}o}, {Brocato}, {Brun}, {Burleigh}, {Burston},
  {Cabrera}, {Cassisi}, {Chaplin}, {Charpinet}, {Chiappini}, {Church},
  {Csizmadia}, {Cunha}, {Damasso}, {Davies}, {Deeg}, {D{\'\i}az}, {Dreizler},
  {Dreyer}, {Eggenberger}, {Ehrenreich}, {Eigm{\"u}ller}, {Erikson}, {Farmer},
  {Feltzing}, {de Oliveira Fialho}, {Figueira}, {Forveille}, {Fridlund},
  {Garc{\'\i}a}, {Giommi}, {Giuffrida}, {Godolt}, {Gomes da Silva}, {Granzer},
  {Grenfell}, {Grotsch-Noels}, {G{\"u}nther}, {Haswell}, {Hatzes},
  {H{\'e}brard}, {Hekker}, {Helled}, {Heng}, {Jenkins}, {Johansen},
  {Khodachenko}, {Kislyakova}, {Kley}, {Kolb}, {Krivova}, {Kupka}, {Lammer},
  {Lanza}, {Lebreton}, {Magrin}, {Marcos-Arenal}, {Marrese}, {Marques},
  {Martins}, {Mathis}, {Mathur}, {Messina}, {Miglio}, {Montalban}, {Montalto},
  {Monteiro}, {Moradi}, {Moravveji}, {Mordasini}, {Morel}, {Mortier},
  {Nascimbeni}, {Nelson}, {Nielsen}, {Noack}, {Norton}, {Ofir}, {Oshagh},
  {Ouazzani}, {P{\'a}pics}, {Parro}, {Petit}, {Plez}, {Poretti}, {Quirrenbach},
  {Ragazzoni}, {Raimondo}, {Rainer}, {Reese}, {Redmer}, {Reffert},
  {Rojas-Ayala}, {Roxburgh}, {Salmon}, {Santerne}, {Schneider}, {Schou},
  {Schuh}, {Schunker}, {Silva-Valio}, {Silvotti}, {Skillen}, {Snellen}, {Sohl},
  {Sousa}, {Sozzetti}, {Stello}, {Strassmeier}, {{\v{S}}vanda}, {Szab{\'o}},
  {Tkachenko}, {Valencia}, {Van Grootel}, {Vauclair}, {Ventura}, {Wagner},
  {Walton}, {Weingrill}, {Werner}, {Wheatley}, \&
  {Zwintz}}]{2014ExA....38..249R}
{Rauer}, H., {Catala}, C., {Aerts}, C., {et~al.} 2014, Experimental Astronomy,
  38, 249

\bibitem[{{Rozner} {et~al.}(2022){Rozner}, {Glanz}, {Perets}, \&
  {Grishin}}]{2022ApJ...931...10R}
{Rozner}, M., {Glanz}, H., {Perets}, H.~B., \& {Grishin}, E. 2022, \apj, 931,
  10

\bibitem[{{Sanchis-Ojeda} {et~al.}(2014){Sanchis-Ojeda}, {Rappaport}, {Winn},
  {Kotson}, {Levine}, \& {El Mellah}}]{2014ApJ...787...47S}
{Sanchis-Ojeda}, R., {Rappaport}, S., {Winn}, J.~N., {et~al.} 2014, \apj, 787,
  47

\bibitem[{{Sarkis} {et~al.}(2021){Sarkis}, {Mordasini}, {Henning}, {Marleau},
  \& {Molli{\`e}re}}]{2021A&A...645A..79S}
{Sarkis}, P., {Mordasini}, C., {Henning}, T., {Marleau}, G.~D., \&
  {Molli{\`e}re}, P. 2021, \aap, 645, A79

\bibitem[{{Schlichting} \& {Mukhopadhyay}(2018)}]{2018SSRv..214...34S}
{Schlichting}, H.~E. \& {Mukhopadhyay}, S. 2018, \ssr, 214, 34

\bibitem[{{Schlichting} {et~al.}(2015){Schlichting}, {Sari}, \&
  {Yalinewich}}]{2015Icar..247...81S}
{Schlichting}, H.~E., {Sari}, R., \& {Yalinewich}, A. 2015, \icarus, 247, 81

\bibitem[{{Scott} \& {Sanders}(2009)}]{2009GeCoA..73.5137S}
{Scott}, E. R.~D. \& {Sanders}, I.~S. 2009, \gca, 73, 5137

\bibitem[{{Szab{\'o}} \& {K{\'a}lm{\'a}n}(2019)}]{2019MNRAS.485L.116S}
{Szab{\'o}}, G.~M. \& {K{\'a}lm{\'a}n}, S. 2019, \mnras, 485, L116

\bibitem[{{Szab{\'o}} \& {Kiss}(2011)}]{2011ApJ...727L..44S}
{Szab{\'o}}, G.~M. \& {Kiss}, L.~L. 2011, \apjl, 727, L44

\bibitem[{{Taylor}(2005)}]{topcat_2005}
{Taylor}, M.~B. 2005, in Astronomical Society of the Pacific Conference Series,
  Vol. 347, Astronomical Data Analysis Software and Systems XIV, ed.
  P.~{Shopbell}, M.~{Britton}, \& R.~{Ebert}, 29

\bibitem[{{Thorngren} \& {Fortney}(2018)}]{2018AJ....155..214T}
{Thorngren}, D.~P. \& {Fortney}, J.~J. 2018, \aj, 155, 214

\bibitem[{{Tinetti} {et~al.}(2021){Tinetti}, {Eccleston}, {Haswell}, {Lagage},
  {Leconte}, {L{\"u}ftinger}, {Micela}, {Min}, {Pilbratt}, {Puig}, {Swain},
  {Testi}, {Turrini}, {Vandenbussche}, {Rosa Zapatero Osorio}, {Aret},
  {Beaulieu}, {Buchhave}, {Ferus}, {Griffin}, {Guedel}, {Hartogh}, {Machado},
  {Malaguti}, {Pall{\'e}}, {Rataj}, {Ray}, {Ribas}, {Szab{\'o}}, {Tan},
  {Werner}, {Ratti}, {Scharmberg}, {Salvignol}, {Boudin}, {Halain}, {Haag},
  {Crouzet}, {Kohley}, {Symonds}, {Renk}, {Caldwell}, {Abreu}, {Alonso},
  {Amiaux}, {Berth{\'e}}, {Bishop}, {Bowles}, {Carmona}, {Coffey},
  {Colom{\'e}}, {Crook}, {D{\'e}sjonqueres}, {D{\'\i}az}, {Drummond},
  {Focardi}, {G{\'o}mez}, {Holmes}, {Krijger}, {Kovacs}, {Hunt}, {Machado},
  {Morgante}, {Ollivier}, {Ottensamer}, {Pace}, {Pagano}, {Pascale}, {Pearson},
  {M{\o}ller Pedersen}, {Pniel}, {Roose}, {Savini}, {Stamper}, {Szirovicza},
  {Szoke}, {Tosh}, {Vilardell}, {Barstow}, {Borsato}, {Casewell}, {Changeat},
  {Charnay}, {Civi{\v{s}}}, {Coud{\'e} du Foresto}, {Coustenis}, {Cowan},
  {Danielski}, {Demangeon}, {Drossart}, {Edwards}, {Gilli}, {Encrenaz}, {Kiss},
  {Kokori}, {Ikoma}, {Morales}, {Mendon{\c{c}}a}, {Moneti}, {Mugnai},
  {Garc{\'\i}a Mu{\~n}oz}, {Helled}, {Kama}, {Miguel}, {Nikolaou}, {Pagano},
  {Panic}, {Rengel}, {Rickman}, {Rocchetto}, {Sarkar}, {Selsis}, {Tennyson},
  {Tsiaras}, {Venot}, {Vida}, {Waldmann}, {Yurchenko}, {Szab{\'o}}, {Zellem},
  {Al-Refaie}, {Perez Alvarez}, {Anisman}, {Arhancet}, {Ateca}, {Baeyens},
  {Barnes}, {Bell}, {Benatti}, {Biazzo}, {B{\l}{\k{e}}cka}, {Bonomo}, {Bosch},
  {Bossini}, {Bourgalais}, {Brienza}, {Brucalassi}, {Bruno}, {Caines},
  {Calcutt}, {Campante}, {Canestrari}, {Cann}, {Casali}, {Casas}, {Cassone},
  {Cara}, {Carmona}, {Carone}, {Carrasco}, {Changeat}, {Chioetto},
  {Cortecchia}, {Czupalla}, {Chubb}, {Ciaravella}, {Claret}, {Claudi},
  {Codella}, {Garcia Comas}, {Cracchiolo}, {Cubillos}, {Da Peppo}, {Decin},
  {Dejabrun}, {Delgado-Mena}, {Di Giorgio}, {Diolaiti}, {Dorn}, {Doublier},
  {Doumayrou}, {Dransfield}, {Dumaye}, {Dunford}, {Jimenez Escobar}, {Van
  Eylen}, {Farina}, {Fedele}, {Fern{\'a}ndez}, {Fleury}, {Fonte}, {Fontignie},
  {Fossati}, {Funke}, {Galy}, {Garai}, {Garc{\'\i}a}, {Garc{\'\i}a-Rigo},
  {Garufi}, {Germano Sacco}, {Giacobbe}, {G{\'o}mez}, {Gonzalez},
  {Gonzalez-Galindo}, {Grassi}, {Griffith}, {Guarcello}, {Goujon}, {Gressier},
  {Grzegorczyk}, {Guillot}, {Guilluy}, {Hargrave}, {Hellin}, {Herrero},
  {Hills}, {Horeau}, {Ito}, {Jessen}, {Kabath}, {K{\'a}lm{\'a}n}, {Kawashima},
  {Kimura}, {Kn{\'\i}{\v{z}}ek}, {Kreidberg}, {Kruid}, {Kruijssen},
  {Kubel{\'\i}k}, {Lara}, {Lebonnois}, {Lee}, {Lefevre}, {Lichtenberg},
  {Locci}, {Lombini}, {Sanchez Lopez}, {Lorenzani}, {MacDonald}, {Magrini},
  {Maldonado}, {Marcq}, {Migliorini}, {Modirrousta-Galian}, {Molaverdikhani},
  {Molinari}, {Molli{\`e}re}, {Moreau}, {Morello}, {Morinaud}, {Morvan},
  {Moses}, {Mouzali}, {Nakhjiri}, {Naponiello}, {Narita}, {Nascimbeni},
  {Nikolaou}, {Noce}, {Oliva}, {Palladino}, {Papageorgiou}, {Parmentier},
  {Peres}, {P{\'e}rez}, {Perez-Hoyos}, {Perger}, {Cecchi Pestellini},
  {Petralia}, {Philippon}, {Piccialli}, {Pignatari}, {Piotto}, {Podio},
  {Polenta}, {Preti}, {Pribulla}, {Lopez Puertas}, {Rainer}, {Reess}, {Rimmer},
  {Robert}, {Rosich}, {Rossi}, {Rust}, {Saleh}, {Sanna}, {Schisano},
  {Schreiber}, {Schwartz}, {Scippa}, {Seli}, {Shibata}, {Simpson}, {Shorttle},
  {Skaf}, {Skup}, {Sobiecki}, {Sousa}, {Sozzetti}, {{\v{S}}poner}, {Steiger},
  {Tanga}, {Tackley}, {Taylor}, {Tecza}, {Terenzi}, {Tremblin}, {Tozzi},
  {Triaud}, {Trompet}, {Tsai}, {Tsantaki}, {Valencia}, {Carine Vandaele}, {Van
  der Swaelmen}, {Adibekyan}, {Vasisht}, {Vazan}, {Del Vecchio}, {Waltham},
  {Wawer}, {Widemann}, {Wolkenberg}, {Hou Yip}, {Yung}, {Zilinskas},
  {Zingales}, \& {Zuppella}}]{2021arXiv210404824T}
{Tinetti}, G., {Eccleston}, P., {Haswell}, C., {et~al.} 2021, arXiv e-prints,
  arXiv:2104.04824

\bibitem[{{Traub}(2011)}]{2011ESS.....2.2403T}
{Traub}, W.~A. 2011, in AAS/Division for Extreme Solar Systems Abstracts,
  Vol.~2, AAS/Division for Extreme Solar Systems Abstracts, 24.03

\bibitem[{{Wilson} {et~al.}(2019){Wilson}, {Hearty}, {Skrutskie}, {Majewski},
  {Holtzman}, {Eisenstein}, {Gunn}, {Blank}, {Henderson}, {Smee}, {Nelson},
  {Nidever}, {Arns}, {Barkhouser}, {Barr}, {Beland}, {Bershady}, {Blanton},
  {Brunner}, {Burton}, {Carey}, {Carr}, {Colque}, {Crane}, {Damke}, {Davidson},
  {Dean}, {Di Mille}, {Don}, {Ebelke}, {Evans}, {Fitzgerald}, {Gillespie},
  {Hall}, {Harding}, {Harding}, {Hammond}, {Hancock}, {Harrison}, {Hope},
  {Horne}, {Karakla}, {Lam}, {Leger}, {MacDonald}, {Maseman}, {Matsunari},
  {Melton}, {Mitcheltree}, {O'Brien}, {O'Connell}, {Patten}, {Richardson},
  {Rieke}, {Rieke}, {Roman-Lopes}, {Schiavon}, {Sobeck}, {Stolberg}, {Stoll},
  {Tembe}, {Trujillo}, {Uomoto}, {Vernieri}, {Walker}, {Weinberg}, {Young},
  {Anthony-Brumfield}, {Bizyaev}, {Breslauer}, {De Lee}, {Downey}, {Halverson},
  {Huehnerhoff}, {Klaene}, {Leon}, {Long}, {Mahadevan}, {Malanushenko},
  {Nguyen}, {Owen}, {S{\'a}nchez-Gallego}, {Sayres}, {Shane}, {Shectman},
  {Shetrone}, {Skinner}, {Stauffer}, \& {Zhao}}]{wilson_2019}
{Wilson}, J.~C., {Hearty}, F.~R., {Skrutskie}, M.~F., {et~al.} 2019, \pasp,
  131, 055001

\bibitem[{{Winn} {et~al.}(2010){Winn}, {Fabrycky}, {Albrecht}, \&
  {Johnson}}]{2010ApJ...718L.145W}
{Winn}, J.~N., {Fabrycky}, D., {Albrecht}, S., \& {Johnson}, J.~A. 2010, \apjl,
  718, L145

\bibitem[{{Winn} \& {Fabrycky}(2015)}]{2015ARA&A..53..409W}
{Winn}, J.~N. \& {Fabrycky}, D.~C. 2015, \araa, 53, 409

\bibitem[{{Wu}(2019)}]{wu}
{Wu}, Y. 2019, \apj, 874, 91

\bibitem[{{Zasowski} {et~al.}(2017){Zasowski}, {Cohen}, {Chojnowski},
  {Santana}, {Oelkers}, {Andrews}, {Beaton}, {Bender}, {Bird}, {Bovy},
  {Carlberg}, {Covey}, {Cunha}, {Dell'Agli}, {Fleming}, {Frinchaboy},
  {Garc{\'\i}a-Hern{\'a}ndez}, {Harding}, {Holtzman}, {Johnson}, {Kollmeier},
  {Majewski}, {M{\'e}sz{\'a}ros}, {Munn}, {Mu{\~n}oz}, {Ness}, {Nidever},
  {Poleski}, {Rom{\'a}n-Z{\'u}{\~n}iga}, {Shetrone}, {Simon}, {Smith},
  {Sobeck}, {Stringfellow}, {Szigeti{\'a}ros}, {Tayar}, \&
  {Troup}}]{zasowski_2017}
{Zasowski}, G., {Cohen}, R.~E., {Chojnowski}, S.~D., {et~al.} 2017, \aj, 154,
  198

\bibitem[{{Zhou} {et~al.}(2019){Zhou}, {Huang}, {Bakos}, {Hartman}, {Latham},
  {Quinn}, {Collins}, {Winn}, {Wong}, {Kov{\'a}cs}, {Csubry}, {Bhatti},
  {Penev}, {Bieryla}, {Esquerdo}, {Berlind}, {Calkins}, {de Val-Borro},
  {Noyes}, {L{\'a}z{\'a}r}, {Papp}, {S{\'a}ri}, {Kov{\'a}cs}, {Buchhave},
  {Szklenar}, {B{\'e}ky}, {Johnson}, {Cochran}, {Kniazev}, {Stassun}, {Fulton},
  {Shporer}, {Espinoza}, {Bayliss}, {Everett}, {Howell}, {Hellier}, {Anderson},
  {Collier Cameron}, {West}, {Brown}, {Schanche}, {Barkaoui}, {Pozuelos},
  {Gillon}, {Jehin}, {Benkhaldoun}, {Daassou}, {Ricker}, {Vanderspek},
  {Seager}, {Jenkins}, {Lissauer}, {Armstrong}, {Collins}, {Gan}, {Hart},
  {Horne}, {Kielkopf}, {Nielsen}, {Nishiumi}, {Narita}, {Palle}, {Relles},
  {Sefako}, {Tan}, {Davies}, {Goeke}, {Guerrero}, {Haworth}, \&
  {Villanueva}}]{2019AJ....158..141Z}
{Zhou}, G., {Huang}, C.~X., {Bakos}, G.~{\'A}., {et~al.} 2019, \aj, 158, 141

\end{thebibliography}

\end{document}